\documentclass[letterpaper]{elsart}
\usepackage{lineno,hyperref}
\usepackage{amsmath}
\usepackage{amsfonts}
\usepackage{bm}
\usepackage{color}
\usepackage{colortbl}
\usepackage{dlfltxbcodetips}
\usepackage{multirow}
\usepackage{setspace}
\usepackage{subfigure}
\usepackage{verbatim}

\newcommand{\specialcell}[2][c]{%
 \begin{tabular}[#1]{@{}c@{}}#2\end{tabular}}

\newcommand \D [2]{\frac{\partial #1}{\partial #2}}

\renewcommand{\vec}[1]{\bm{\mathrm{#1}}}

\def \F{\vec{F}}

\def \g{\vec{g}}
\def \G{\vec{G}}

\def \X{\vec{X}}

\def \cG{\mathcal{G}}

\def \s{\vec{s}}
\def \u{\vec{u}}

\def \x{\vec{x}}

\def \div{\nabla \cdot \mbox{}}
\def \grad{\nabla}
\def \lap{\nabla^2}

\def \Ds{{\mathrm d}\s}
\def \Dx{{\mathrm d}\x}

\begin{document}

\begin{frontmatter}

%\mbox{}

%\vspace{-\baselineskip}

\title{A fully resolved active musculo-mechanical model for esophageal transport}

\author{Wenjun Kou}
\address{Theoretical and Applied Mechanics, Northwestern
 University, 2145 Sheridan Road, Evanston, Illinois 60208, USA}
 
\author{Amneet Pal Singh Bhalla}
\address{Department of Mechanical Engineering, Northwestern
 University, 2145 Sheridan Road, Evanston, Illinois 60208, USA}
\address{Courant Institute of Mathematical Sciences, New York University,
 New York, NY 10012, USA}
 
\author{Boyce E.~Griffith}
\address{Department of Mathematics, University of North Carolina at Chapel Hill, Phillips Hall, Campus Box 3250, Chapel Hill, North Carolina 27599-3250, USA}

%\address{Leon H.~Charney Division of Cardiology, Department of
% Medicine, New York University School of Medicine, 550 First Avenue,
% New York, New York 10016, USA}
\author{John E.~Pandolfino}
\address{Department of Medicine, Feinberg School of Medicine, Northwestern University, 676 North Saint Clair Street, 14th Floor, Chicago, Illinois 60611, USA}
\author{ Peter J.~Kahrilas}
\address{Department of Medicine, Feinberg School of Medicine, Northwestern University, 676 North Saint Clair Street, 14th Floor, Chicago, Illinois 60611, USA}
\author{Neelesh A.~Patankar} \ead{n-patankar@northwestern.edu}
\address{Department of Mechanical Engineering, Northwestern
 University, 2145 Sheridan Road, Evanston, Illinois 60208, USA}

\begin{abstract}
Esophageal transport is a physiological process that mechanically
    transports an ingested food bolus from the pharynx to the stomach
    via the esophagus, a multi-layered muscular tube.  This process
    involves interactions between the bolus, the esophagus, and the
    neurally coordinated activation of the esophageal muscles.  In
    this work, we use an immersed boundary (IB) approach to simulate
    peristaltic transport in the esophagus.  The bolus is treated as a
    viscous fluid that is actively transported by the muscular
    esophagus, which is modeled as an actively contracting,
    fiber-reinforced tube.  A simplified version of our model is
    verified by comparison to an analytic solution to the tube
    dilation problem.  Three different complex models of the
    multi-layered esophagus, which differ in their activation patterns
    and the layouts of the mucosal layers, are then extensively
    tested.  To our knowledge, these simulations are the first of
    their kind to incorporate the bolus, the multi-layered esophagus
    tube, and muscle activation into an integrated model.  Consistent with experimental observations, our
    simulations capture the pressure peak generated by the muscle
    activation pulse that travels along the bolus tail. These fully
    resolved simulations provide new insights into roles of the
    mucosal layers during bolus transport. In addition, the information on pressure and the kinematics of the esophageal wall due to the coordination of muscle activation is provided, which may help relate clinical data from manometry and ultrasound images to the underlying esophageal motor function.
\end{abstract}

\begin{keyword}
    fluid-structure interaction \sep immersed boundary method \sep   esophageal transport \sep muscle activation
\end{keyword}

\end{frontmatter}

\section{Introduction}

Interactions between fluids and deformable structures are widespread
in biological systems, and such interactions often involve complex
moving interfaces and large structural
deformations~\cite{AnRFM2011}. Esophageal transport is one such
process, whereby the food bolus is transported to the stomach via the
esophagus.  The esophagus is a flexible, multi-layered tube that
consists of mucosal, interfacial, circumferential, and longitudinal
muscle layers.  The pumping force required to produce this peristaltic
transport process is generated by neurally coordinated muscle
activation along the esophagus~\cite{Kahrilas1997,Mittal2006}, and
accounting for the full physiological details of the transport process
is challenging.  Simplified analytical models can provide some
insights into this biophysical process~\cite{Li-Brasseur1993} but are
often limited in their scope.  More complete models are needed to
investigate esophageal pathophysiology, such as motility disorders,
and hold the potential to advance diagnoses and patient treatment.

Current studies on the modeling of esophageal transport have focused
on specific, albeit important, subproblems, such as characterizing the material properties of each layer of the
esophagus tube~\cite{Fan2004,Natali2009,Sokolis2013,Yang2006a,Yang2006b,Stavropoulou2009},
investigating the flow of bolus with specified time-dependent
bolus geometry or known lumen
pressure~\cite{Li-Brasseur1993,Li1994,Ghosh2005}, and estimating the muscle active tension based on known time-dependent pressure
distribution~\cite{Ghosh2008,Nicosia-Brasseur2002}.  To the best of
our knowledge, however, there is presently no computational model of
esophageal transport that couples models of the bolus, the esophageal
structure, and the muscle activation within an integrative numerical
model.

This work presents one such integrative model of esophagael transport
that is based on the immersed boundary (IB) method \cite{CSPeskin02}.
The IB method is an approach to modeling fluid-structure interaction
that was introduced to simulate the fluid dynamics of heart
valves~\cite{Peskin1972,Peskin1977}, and which has subsequently been
applied to a broad range of problems in biology \cite{CSPeskin02}.
The IB method uses an Eulerian description of the momentum and
incompressibility of the coupled fluid-structure system along with a
Lagrangian description of the structural forces produced by the
elasticity or active tension generation of the structure.  The primary
advantage of this formulation is that it avoids the need to employ
body-fitted grids, and thereby eliminates the need to develop complex
remeshing strategies as the immersed structure
deforms~\cite{CSPeskin02,Mittal&Iaccarino2005}. In this paper, we
present a fully resolved fluid-structure intreaction model of
esophageal transport, which includes detailed descriptions of the
esophageal wall, the bolus, and their interaction.  The model is fully
resolved in the sense that it does not assume simplified fluid
dynamics or structural deformations.

In our model, the majority of the computational domain is occupied by
the immersed body, with a fluid (i.e., the bolus) confined in a narrow
lumen.  Thus, it is important to describe the mechanical response of
the esophagus tube.  To that end, we discretize the continuous fibers
of the esophagus into springs and beams and associate a volumetric
patch with each such spring and beam. This allows us to compute the
spring and beam parameters from the material properties of the
esophagus (e.g., its Young's modulus). To test the fiber-based
esophagus tube model, we simulate the problem of dilation of a
three-dimensional tube and compare the numerically obtained inner
fluid pressure with an analytically derived solution.

To simulate bolus transport in a physiologically realistic manner, an
esophagus model is constructed as a four-layered structure. Muscle
activation that results in the peristaltic motion of the esophagus is
modeled via springs with dynamic rest lengths. To handle the numerical
challenge arising from large deformations of the mucosal layer, we
employ a locally refined structural discretization that ensures that
the Lagrangian structure does not ``leak'' even under very large
deformations~\cite{CSPeskin02}. We consider three cases with different
muscle activation models and mucosal layer fiber arrangements, and we
discuss the key features related to muscle cross-section area and
pressure peaks during bolus transport.

\section{Mathematical formulation}
\subsection{The immersed boundary method}
The IB formulation of problems of fluid-solid interaction employs an
Eulerian description for the momentum equation and the divergence-free
condition and a Lagrangian description of the deformation of the
immersed structure and the resulting structural forces. Here, we use
the same notation for the Eulerian and Lagrangian coordinates as
detailed in Griffith~\cite{Griffith2012}. Specifically, we let $\x =
(x_1, x_2, x_3) \subset \Omega$ denote fixed Cartesian coordinates, in
which $\Omega \subset \mathbb{R}^3 $ denotes the fixed domain occupied
by the entire fluid-structure system. We use $\s = (s_1, s_2, s_3)
\subset U$ to denote the Lagrangian coordinates attached to the
immersed structure, in which $U \subset \mathbb{R}^3$ denotes the
fixed material coordinate system attached to the structure. For
simplicity of implementation, we consider that the fluid-structure
system possesses a uniform mass density $\rho $ and dynamic viscosity
$\mu$. This simplification implies that the immersed structure is
neutrally buoyant and viscoelastic rather than purely elastic. An
extension of the present mathematical formulation to problems with
nonuniform mass densities or viscosities is also feasible; see~Ref.~\cite{Fai2013} for details.

The equations of motion of the coupled fluid-structure system are \cite{CSPeskin02}
\begin{align}
 \MoveEqLeft[4] \rho\left(\D{\u}{t}(\x,t) + \u(\x,t) \cdot \grad \u(\x,t) \right) = -\grad p(\x,t) + \mu \lap \u(\x,t) \nonumber \\
 & \quad \quad \quad \mbox{} + \g(\x,t), \label{eqn_momentum}\\
 \div \u(\x,t) &= 0, \label{eqn_continuity} \\
 \g(\x,t) &= \int_{\Omega} \G(\s,t) \, \delta(\x - \X(\s,t)) \, \Ds, \label{eqn_F_f} \\
 \D{\X}{t} (\s,t) &= \int_{U} \u(\x,t) \, \delta(\x - \X(\s,t)) \, \Dx. \label{eqn_u_interpolation} \\
 \G(\s,t) &= \cG[\X(\s,t)]. \label{eqn_elastForce} 
\end{align}
Eqs.~\eqref{eqn_momentum} and \eqref{eqn_continuity} are the incompressible 
Navier-Stokes equations written in the Eulerian form,  $\u(\x,t)$ is the Eulerian 
velocity, $p(\x,t)$ is the pressure, and $\g(\x,t)$ is the Eulerian elastic force density. 
Eq.~\eqref{eqn_elastForce} describes the elastic force in the immersed body in  
Lagrangian form, in which $\G(\s,t)$ is the elastic force density and  
$ \cG: \X \mapsto \G$ is a
time-dependent functional that determines the Lagrangian force density
from the current configuration of the immersed structure.
Interactions between Lagrangian and Eulerian variables in
eqs.~\eqref{eqn_F_f} and \eqref{eqn_u_interpolation} are mediated by
integral transforms with a three-dimensional Dirac delta function
kernel $\delta(\x) = \Pi_{i=1}^{3}\delta(x_i)$. Specifically,
eq.~\eqref{eqn_F_f} converts the Lagrangian force density $\G(\s,t)$
into an equivalent Eulerian force density $\g(\x,t)$, and
eq.~\eqref{eqn_u_interpolation} determines the physical velocity of
each Lagrangian material point from the Eulerian velocity
field, thereby effectively imposing the no-slip condition along the
fluid-solid interface. The discretized versions of these equations used in this work employ a
regularized version of the delta function, denoted $\delta_h(\x) =
\Pi_{i=1}^{3}\delta_h(x_i)$; for details on the construction of such
regularized delta functions, see Ref.~\cite{CSPeskin02}. The details
on the spatial discretization of Eulerian fluid system (i.e.,
eqs.~\eqref{eqn_momentum} and \eqref{eqn_continuity}),
Lagrangian-Eulerian interaction equations, (i.e., eqs. \eqref{eqn_F_f}
and \eqref{eqn_u_interpolation}), and the temporal discretization of
the system of equations can be found
in~refs.~\cite{Griffith2012,Griffith2009}.  In the following section,
we discuss the Lagrangian discretization of eq.~\eqref{eqn_elastForce}
to characterize the material elasticity of the immersed structure.

% a function that computes the elastic force density. The interactions between the Lagrangian and Eulerian variables are mediated through a three-dimensional Dirac delta function $\delta(\x) = \Pi_{i=1}^{3}\delta(x_i)$ as described in eqs.~\eqref{eqn_F_f} and\eqref{eqn_u_interpolation}. In particular eq.~\eqref{eqn_F_f} converts the Lagrangian force density $\G(\s,t)$ into an equivalent Eulerian force density $\g(\x,t)$. Eq.~\eqref{eqn_u_interpolation} updates the physical position of each Lagrangian material point based on an interpolated Eulerian velocity field; thereby imposing a no-slip boundary condition on the fluid-structure interface implicitly. For details on the construction of Dirac delta function and properties of eqs.~\eqref{eqn_F_f} and \eqref{eqn_u_interpolation}, we refer readers to~\cite{CSPeskin02}. The details on the spatial discretization of Eulerian fluid system (i.e., eqs.~\eqref{eqn_momentum} and \eqref{eqn_continuity}), Lagrangian-Eulerian interaction equations, (i.e., eqs. \eqref{eqn_F_f} and \eqref{eqn_u_interpolation}) and the temporal discretization of the system of equations can be found in~\cite{Griffith2012,Griffith2009}. In the following section we discuss the Lagrangian discretization of eq.~\eqref{eqn_elastForce} in order to characterize the material elasticity of the immersed structure.
 
%%\section{Material elasticity} 
\subsection{Material elasticity}
The specific form of the mapping function $ \cG: \X \mapsto \g$ is dictated by the model of 
material elasticity of the immersed body. To that end, Chadwick~\cite{Chadwick1982}, Ohayon and Chadwick~\cite{Ohayon&Chadwick1988}, and Tozeren~\cite{Tozeren1985} proposed 
the ``fluid-fiber" and the ``fluid-fiber-collagen" models to characterize the material 
elasticity of biological tissues. They assumed the tissues to be an aggregation of elastic 
fibers that are embedded in a soft matrix. These models were used to describe 
the esophageal wall by Nicosia and Brasseur~\cite{Nicosia-Brasseur2002}. They 
considered the muscle layer as a family of fibers and discarded the elasticity of the soft 
matrix. Inspired by the success of their model~\cite{Nicosia-Brasseur2002}, we also ignore the 
elasticity of the soft ground matrix in the esophageal wall, and model all esophageal layers as families of continuous fibers embedded in the background fluid. Thus, the material elasticity of the esophagus tube is essentially 
represented by the fiber elasticity, which is approximated using springs and beams in our discretized IB scheme. By contrast, Ghosh et al.~\cite{Ghosh2008} have 
modeled the esophageal muscle layer as a family of incompressible continuous fibers embedded 
in a soft isotropic matrix. Such models will be explored in our future work.

We describe the fiber-based material elasticity in terms of a strain-energy functional 
$E=E[\X(\cdot,t)]$. The corresponding Lagrangian elastic force density can be 
derived by taking the Frechet derivative of $ E $ as 
\begin{equation}
\wp E[\X(\cdot, t)] = - \int_{U} \G(\s, t) \cdot \wp \X(\s,t)\, \Ds,
\end{equation}
in which $\wp$ denotes the perturbation of a quantity. Since the elastic properties of the structure 
are described in terms of a family of elastic fibers that resist extension, compression and 
bending, the strain-energy functional $E=E[\X(\cdot,t)]$ can be decomposed into a stretching 
part $ E_s$ that accounts for the extension and compression of the fibers, and a bending 
part $E_b$ which accounts for the resistance of the fibers to bending, i.e.,
$E = E_s + E_b$.

\subsection{Elastic springs and beams} \label{sec_springs_beams}

We use a 3D spring network to represent the elastic stretching energy $E_s$ of the esophagus tube.
Three families of springs are employed for the radial, circumferential, and axial fibers 
of the tube. On the contrary, we only include axial beams in esophagus model to account for the bending energy $E_b$, as the significant curvature change for the long esophageal tube occurs primarily along the axial direction. 

In our implementation of the IB method \cite{Griffith2012}, instead of computing the elastic force 
density, the total elastic force is computed for a Lagrangian node, which is then spread to the 
Eulerian grid (via eq.~\eqref{eqn_F_f}) to obtain the equivalent Eulerian force density. To relate 
the spring/beam constant with the material property of the esophagus (such as its Young's modulus), we associate a 
volumetric patch with each spring and beam. To obtain the volumetric patch, we describe the tube in 
cylindrical coordinates $(r, \theta, z)$ with the reference configuration given 
by $(a \leq r \leq b, 0 \leq \theta \leq 2\pi, 0 \leq z \leq l )$, where $a$, $b$ and $l$ are the inner radius, the outer radius and the length of the tube in the reference configuration, respectively. Let $(r_i, \theta_j, z_k)$ 
denote the node point of the tube, with $ \Delta r_i = r_{i+1} -r_i$, $\Delta \theta_j = \theta_{j+1} - \theta_j $ and 
$\Delta z_k = z_{k+1} - z_k$ denoting the spacing 
along $r, \theta$ and $z$ coordinates, respectively. Then, a case of uniform spacing 
$\Delta r,\; \Delta \theta $ and $\Delta z$ will be 
obtained if we let $\Delta r_i=\Delta r,\; \Delta \theta_j = \Delta \theta$ and $\Delta z_k=\Delta z$ 
for the discretization. Depending upon whether any of the three types of springs or an axial beam lies in the 
interior of the structure or its boundary, different volumetric patches (interior or boundary patches) 
will be associated with them.  We layout the nodal points in such a way that volumetric patches of each type  $P_{\text{type}} $ 
(i.e., patches associated with radial/axial/circumferential springs or axial beams) do not overlap with
each other and they sum up to the volume of the esophagus tube. This can be written as   
\begin{equation}
 \begin{cases}
 P^I_{\text{type}} \bigcap P^J_{\text{type}} = \varnothing, & I \neq J, \\
 \bigcup_{I}P^I_{\text{type}} = \sum_{I} P^{I}_{\text{type}} = V_{\text{esophagus}} & I \in N_{\text{type}},\\ 
 \end{cases} \label{eqn_PatchS}
 \end{equation}
in which, $ V_{\text{esophagus}} $ is the volume of the esophagus and $N_{\text{type}}$ is the number of patches 
of the same type.

\begin{figure}
 \centering
 \subfigure[]{
 \includegraphics[scale = 0.33]{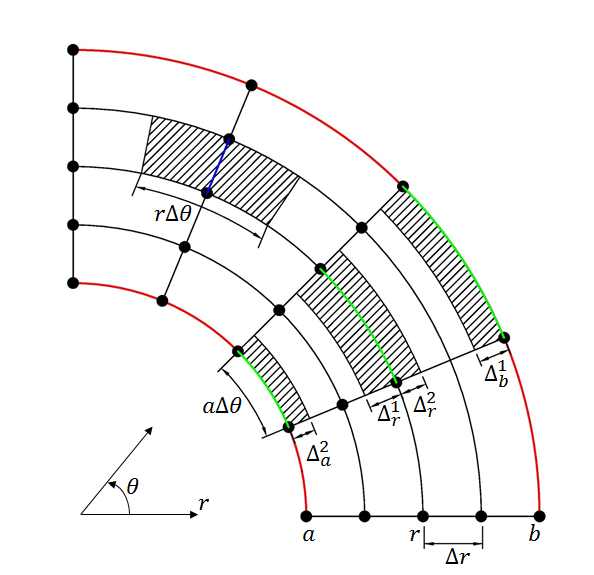} 
 \label{fig:patch_rt_uniform}
 }
 \subfigure[]{
 \includegraphics[scale = 0.30]{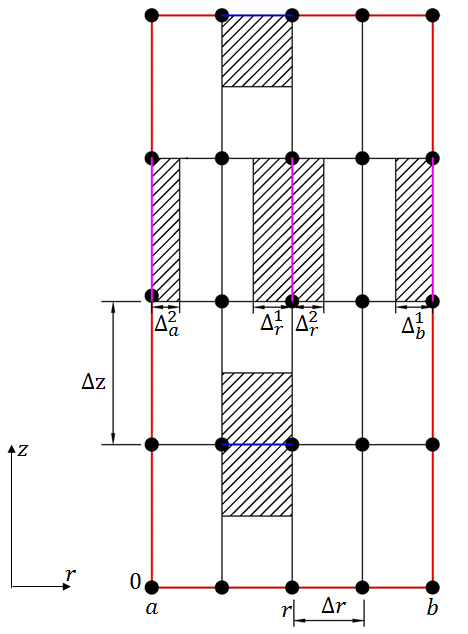}
 \label{fig:patch_rz}
 } 
 \caption{\subref{fig:patch_rt_uniform} Schematic of patches (hatched areas) associated with 
 radial springs (blue lines) and circumferential springs (green curves) in $(r, \theta)$ plane. The 
 red curves denote the boundaries of the tube: the inner $(r=a)$ and the outer surface $(r=b)$. In 
 $(r, \theta)$ plane, the patch of a radial spring connected by nodes $(r, \theta, z)$ and 
 $(r+\Delta r,\theta, z)$ is $(r, r+\Delta r) \times (\theta -0.5\Delta \theta, \theta + 0.5\Delta \theta)$. 
 The patch of a circumferential spring connected by the nodes $(r,\theta, z)$ and $(r,\theta+\Delta \theta, z)$ 
 is $(r-\Delta_r^1, r+\Delta_r^2) \times (\theta, \theta+\Delta \theta)$ for $a<r<b$; $(r-\Delta_r^1, r) \times (\theta, \theta+\Delta \theta)$ 
 for $r=b$ and $(r, r+\Delta_r^2) \times (\theta, \theta+\Delta \theta)$ for $r=a$. Here $\Delta_{r}^1 = r - \sqrt{r (r - \Delta r)}$ 
 and $\Delta_{r}^2 = - r + \sqrt{r (r + \Delta r)}$ are such that the circumferential patch is 
 partitioned into two equal volumes.
 \subref{fig:patch_rz} Schematic of patches (hatched areas) associated with radial springs (blue lines) 
 and axial springs (magenta lines) in $(r,z)$ plane. The red curves denote the boundaries of the tube: 
 the inner $(r=a)$, outer $(r=b)$, lower  $(z=0)$ and the upper surface $(z=l)$. In $(r, z)$ plane, 
 the patch of a radial spring connected by nodes $(r, \theta, z)$ and $(r+\Delta r,\theta, z)$ is 
 $(r, r+\Delta r) \times (z-0.5\Delta z, z+0.5\Delta z)$ for $0<z<l$; $(r, r+\Delta r) \times (z-0.5\Delta z, z)$ 
 for $z=l$ and $(r, r+\Delta r) \times (z, z+0.5\Delta z)$ for $z=0$. The patch of an axial spring 
 connected by nodes $(r,\theta, z)$ and $(r,\theta, z+\Delta z)$ is $(r-\Delta_r^1, r+\Delta_r^2) 
 \times (z, z+\Delta z)$ for $a<r<b$; $(r-\Delta_r^1, r) \times (z, z+\Delta z)$ for $r=b$ and  
 $(r, r+\Delta_r^2) \times (z, z+\Delta z)$ for $r=a$.}
 
\end{figure}

\begin{figure}
 \centering
  \subfigure[]{
 \includegraphics[scale = 0.30]{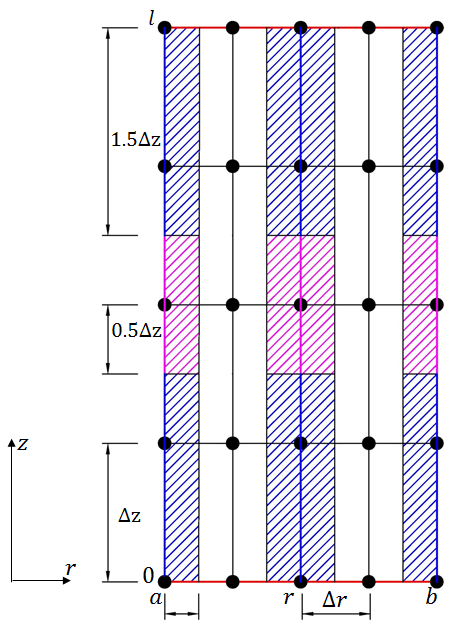} 
 \label{fig-patch_beam_uniform}
 }
 \subfigure[]{
 \includegraphics[scale = 0.30]{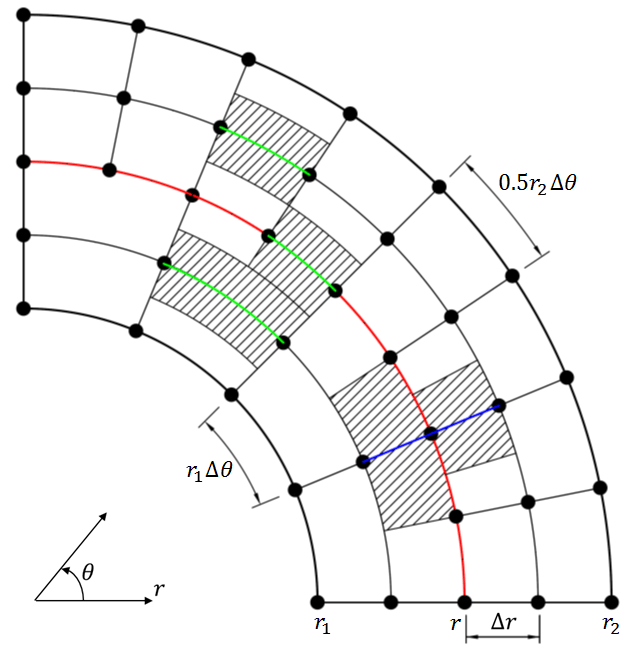}
 \label{fig-patch_nonuniform}
 } 
 \caption{\subref{fig-patch_beam_uniform} Schematic of patches (hatched areas) associated with axial 
 beams in $(r,z)$ plane. The red lines denote the boundaries of the tube: the top  $z=l $ and the 
 bottom $z=0 $. The blue patch is associated with a beam that has a node on the top or bottom 
 surface of the tube and its volume is: $V = 1.5 r_i \Delta z \Delta \theta \Delta r$ for $a < r < b$; or 
 $V = 0.75 r_i \Delta z \Delta \theta \Delta r$, if $(r-a)(r-b)=0$. The magenta patches are associated 
 with a beam which does not have a node at the top or bottom surface of the tube. The associated patch 
 volume is: $V = r_i \Delta z \Delta \theta \Delta r$ for $a <r < b$; or $V = 0.5r_i \Delta z \Delta \theta \Delta r$,
 if $(r-a)(r-b)=0$. 
 \subref{fig-patch_nonuniform} Schematic of patches (hatched areas) associated with 
 radial springs (blue lines) and circumferential springs (green curves) in $(r, \theta)$ plane with 
 nonuniform $\Delta \theta$. The red curve denotes the interface between the two layers with 
 different spacing in $\theta$. Compared with the case of uniform $\Delta \theta$, only circumferential springs 
 located on the interface need special treatment.}
 \label{fig-patch_rt_uniform}
 \end{figure}

Fig.~\ref{fig:patch_rt_uniform} and~\ref{fig:patch_rz} show the volumetric patches of various springs 
in $(r,\theta)$ plane and $(r,z)$ plane, respectively. Fig.~\ref{fig-patch_beam_uniform} shows the 
patches associated with axial beams. Cases with nonuniform $\Delta \theta$ are also considered in 
this work. In particular, we consider cases with $\Delta \theta$ for inner layers and 
$0.5\Delta \theta$ for outer layers of the tube. The patches associated with the circumferential 
springs in $(r, \theta)$ plane for the nonuniform $\Delta \theta$ are shown in 
Fig.~\ref{fig-patch_nonuniform}.  
Once the patches for the discrete springs and beams are obtained, 
the elastic force due to extension, compression and bending of the fibers can be computed with given 
material parameters. Next we discuss how spring/beam constants are obtained in our model from the elastic 
modulus of the fibers.

\emph{Spring constant:} We assume a linear relationship between the stress and strain of a spring, $sp$, 
which provides resistance to extension and compression of a fiber. If the stress and strain of the spring 
are defined with respect to the undeformed configuration, then we have
 \begin{align}
 \sigma &= \frac{F}{A} = S \varepsilon, \\
 \varepsilon &= \frac{l-L} {L},
 \end{align}
in which $\sigma$, $\varepsilon$, and $F$ are the stress, strain, and internal force of the spring, 
respectively, $S$ is the elastic modulus of the fiber, $l$ is the current length, $L$ is the rest 
length and $A$ is the undeformed cross sectional area. Consequently, the 
expression for the spring nodal forces $\F_{sp}^{I_1}$ and $\F_{sp}^{I_2}$ for nodes $I_1$ and 
$I_2$ connected by the spring are given by
 \begin{equation}
 \F_{sp}^{I_1} = - \F_{sp}^{I_2} = \frac{S A}{L} (l-L) \frac{\X_{sp}^{I_2} - \X_{sp}^{I_1}}
 {\left| \X_{sp}^{I_2} - \X_{sp}^{I_1} \right|} = K (l-L) \frac{\X_{sp}^{I_2} - \X_{sp}^{I_1}}
 {\left| \X_{sp}^{I_2} - \X_{sp}^{I_1} \right|}, \label{eqn_sp1}
 \end{equation}
in which, $K = S A /L $ is the spring constant (also called the spring stiffness). The undeformed cross 
sectional area $A$ can be obtained from the associated patch of the spring, $A = V_{sp} / L$, where $V_{sp}$ is the volume of patch of the spring.

\emph{Beam constant:} A beam $b$ associated with three nodes $I_1$, $I_2$, and $I_3$, provides resistance 
to bending and sets a preferred (possibly a time-dependent) curvature of the fiber. The nodal forces for 
the beam $\F_{b}^{I_1}$, $\F_{b}^{I_2} $, and $\F_{b}^{I_3}$ can be obtained from the Frechet 
derivative of the bending energy as
\begin{equation}
E_{b} = \frac{1}{2} \int_{P^{I}_b}c_b \left| \frac{\partial^2 \X_{b}}{\partial s^2} - \frac{\partial^2 \X_{b}^0}{\partial s^2} \right|^2 \text{d}s,
\end{equation}
in which, $c_b$ is the bending stiffness and $P^{I}_b$ is the associated patch of the beam. $\X_{b}^0$ is the preferred configuration, which in present 
work satisfies $\frac{\partial^2 \X_{b}^0}{\partial s^2}=0$. Thus,
\begin{equation}
\int_{P^{I}_b} \G_{b} \cdot \wp \X_b \; \text{d}s = 
- \int_{P^{I}_b} c_b \frac{\partial^2 \X_{b}}{\partial s^2} 
\cdot \frac{\partial^2 \wp \X_{b}}{\partial s^2}\; \text{d}s. \label{eq_beam}
\end{equation}
We approximate the second derivative (with respect to the arc length $s$) in eq.~\eqref{eq_beam} in 
the reference configuration of the tube, where the three beam nodes are assumed to be on the same 
line (i.e., have zero curvature). This basically implies small bending deformation. With $q_i$ (for 
$i=1,2,3$) labeling the three nodes, we construct shape function $N_i$ for node $q_i$ in the local 
coordinate system $q$ to get
$\X_{b} = \sum_{j=1}^3 \X_{b}^{I_j} N_j(q)$ and $\wp \X_{b} = \sum_{k=1}^3 \wp \X_{b}^{I_k} N_k(q)$. 
\begin{align}
N_1(q)&= \frac{(q-q_2)(q-q_3)}{(q_1 -q_2)(q_1-q_3)}; 
 \frac{\partial^2 N_1}{\partial q^2} = \frac{2}{(q_1 -q_2)(q_1-q_3)} \\
N_2(q)&= \frac{(q-q_1)(q-q_3)}{(q_2 -q_1)(q_2-q_3)}; 
 \frac{\partial^2 N_2}{\partial q^2} = \frac{2}{(q_2 -q_1)(q_2-q_3)} \\
N_3(q)&= \frac{(q-q_3)(q-q_1)}{(q_3 -q_1)(q_3-q_2)}; 
 \frac{\partial^2 N_3}{\partial q^2} = \frac{2}{(q_3 -q_1)(q_3-q_2)}
\end{align}
Thus, 
eq.~\eqref{eq_beam} evaluates as
\begin{align}
\int_{P^{I}_b} \G_{b} \cdot \wp \X_{b}\; \text{d}s &= 
- \int_{P^{I}_b} c_b \frac{\partial^2 \X_{b}}{\partial q^2} 
\cdot \frac{\partial^2 \wp \X_{b}}{\partial q^2}\; \text{d}q 
\nonumber \\
&= - V^{I}_b c_b \X_{b}^{I_j} \frac{\partial^2 N_j}{\partial q^2} 
 \cdot \frac{\partial^2 N_k}{\partial q^2} \wp \X_{b}^{I_k}
\nonumber \\
&= \F_{b}^{I_k} \cdot \wp \X_{b}^{I_k},
\end{align}
in which, $V^{I}_b$ is the volume of the patch associated with the beam and 
$\F_{b}^{I_k} = - V^{I}_b c_b \X_{b}^{I_j} \frac{\partial^2 N_j}{\partial q^2} \otimes
\frac{\partial^2 N_k}{\partial q^2}$. For the case of uniform distance between neighboring nodes 
of a beam in the reference configuration, i.e., when $q_2 -q_1 =q_3 -q_2 =\Delta q$,  
simple expression for the nodal forces can be obtained as below,
\begin{equation}
 \left[ \begin{array}{ccc} \F_{b}^{I_1} & \F_{b}^{I_2} & \F_{b}^{I_3} \end{array} \right]^T=
 \frac{-V^{I}_b c_b}{\Delta q^4} \left[ \begin{array}{ccc} 1 
 & -2 & 1 \end{array} \right]^T \otimes
 \left[ \begin{array}{ccc} 1
 & -2 
 & 1 \end{array} \right] 
 \left[ \begin{array}{ccc} \X_{b}^{I_1} & \X_{b}^{I_2} & \X_{b}^{I_3} \end{array} \right]^T.
 \label{eq_beam_force}
\end{equation}
The bending coefficient $c_b$ in the above equation can be determined from the relation $c_b = \frac{S I}{A}$,
in which $I$ and $A$ are the second moment of area and area of cross section of the beam, respectively. 

\section{Verification case: A 3D dilation problem}
\label{sec_tube_dilation}

\subsection{Problem formulation}

For the fiber-based IB scheme, Griffith~\cite{Griffith2009,Griffith2005} has studied convergence properties for various 2D cases. Here we present a test case of a 3D elastic cylindrical tube undergoing dilation to verify the solution 
methodology. A cylindrical tube composed of three families of continuous fibers (i.e., axial, circumferential 
and radial fibers) is immersed in a rectangular fluid domain. We nondimensionalize the system based on the inner radius of 
the cylindrical tube, the density and viscosity of the fluid. The tube's reference configuration is 
described in cylindrical coordinates $(r, \theta, z)$ with $1 \leq r \leq 1+T,\;0 \leq \theta 
\leq 2\pi,\;0 \leq z \leq 10$, in which $T$ is the tube thickness. The fluid domain is described 
in Cartesian coordinates $(x , y , z)$ with $-2 \leq x \leq 2, \; -2 \leq y \leq 2,\; 0 \leq z \leq 10$. The dimensionless Young's modulus 
$S$ of all fibers is taken to be the same, $S = 4 \times 10^4$.

The dilation process includes two phases. The first phase is an inflation phase. 
This is modeled by adding fluid in the domain from the top end while keeping its bottom end closed. The second
phase is the relaxation phase. The relaxation phase continues until a
stationary state is reached, at which the inertial and viscous terms are approximately three orders of magnitude smaller than
the pressure term. Thus, the pressure force from the constraint of fluid incompressibility is 
balanced by the elastic force in the tube. The 
simulations are carried by specifying the following boundary conditions for the fluid domain: traction (normal and tangential) free boundary 
conditions for the four lateral surfaces of the domain; zero-velocity boundary condition for the bottom surface; 
and a time-dependent velocity boundary condition for the top surface. On the inflow surface, the tangential velocity is zero and  the normal velocity is $\u \cdot \vec{n}= -u_0 f(t) (1-x^2-y^2)$ if $x^2+y^2 \leq 1$, and zero otherwise. Here, $f(t)$ is a decreasing function of time which vanishes at the end 
of the inflation process. 
%We also fix the top and bottom ends of the cylindrical tube in place to ensure the immersed structure does not move out of the fluid domain.
 
We assume a plain strain state for this relatively long tube at the final equilibrium state.
Let $U(r)$ denote the radial displacement field of the tube in the middle region, then by assuming plane strain conditions (i.e., a sufficiently long tube) in the final equilibrium state, an analytic expression for the inner pressure can be obtained (see Appendix A),
\begin{eqnarray}
 P_\text{inner} &=&S\left[\log \left(r_\text{i} + \sqrt{r_\text{i}^2 -C}\right) - \frac{\sqrt{r_\text{i}^2 - C}}{r_\text{i}} \right]\\ \nonumber 
 & & - S\left[\log\left(r_\text{o}  + \sqrt{r_\text{o}^2 -C}\right) + \frac{\sqrt{r_\text{o}^2 - C}}{r_\text{o}} - \log\left(\frac{r_\text{i}}{r_\text{o}}\right)\right]. \label{eqn_P_Ur}
\end{eqnarray}
Here, $C = r_\text{i}^2 - R_\text{i}^2 $,  $r_\text{i}$ and $r_\text{o}$ are the deformed inner and outer radius, respectively; $R_\text{i}$ and $R_\text{o}$ denote the initial inner and outer radius, respectively. Thus, $r_\text{i} = R_\text{i} + U(r_\text{i})$, with $U(r_\text{i})$ denoting the radial displacement of the inner surface of the tube and $r_\text{o} = \sqrt{R_\text{o}^2 - R_\text{i}^2 + r_\text{i}^2} $. We use the observed $U(r_\text{i})$ in our simulations, and compare the numerical and predicted analytic values of the inner pressure,  respectively $P_\text{numerical}$ and $P_\text{analytic}$.

\subsection{Verification results}

Here, we conduct test cases with different tube thickness and dilation levels, where higher dilation level is simulated by adding more fluid into the tube during the dilation phase. For each case, the Eulerian computational domain is discretized using an $N \times N \times 50$ Cartesian grid, whereas the Lagrangian structural domain is described using a cylindrical coordinate system $(r, \theta, z)$ and discretized using an $N_r \times N_\theta \times 100$ mesh. To prevent the fluid from leaking out of the structure, we keep the grid size of Lagrangian mesh smaller than that of Eulerian mesh. The grid number for cases with different tube thickness is listed in Table~\ref{table_thickness}. For a thicker tube of thickness $T =1$, we use nonuniform $\Delta \theta$ for inner and outer layers, with $N_\theta=64$ for the inner and $N_\theta=128$ for the outer layers.

\begin{table}[h]
 \caption{Error in the inner pressure for different tube thickness $T$. $U(r_\text{i})$ and $P_{\text{numerical}}$ are measured radial displacement of the inner surface and inner pressure in the middle section of the tube, respectively. The relative error $\epsilon_p = \frac{|P_{\text{numerical}} - P_\text{analytic}|}{P_\text{analytic}}$}
 \centering
\begin{tabular}{l | l | l | l | l | l | l}
 \hline \hline
 $T$ & $N_r \times N_\theta$ & $N$ & $U(r_\text{i})$ & $P_{\text{numerical}}$ & $ P_\text{analytic}$ & $\epsilon_p$\\ [1ex]
 \hline
 0.2 & $4 \times 128$ & $40$ & $ 0.07989 $ & 399.4 & 403.14 & 9.28e-3\\
 0.5 & $10 \times 128$ & $40 $ & $0.07410 $ & 699.9 & 710.78 & 1.53e-2\\
 1 & $6 \times 64 + 4 \times 128 $ & $20$ & 0.07313 & 961.4 & 972.30 &1.12e-2\\
 \hline
 \end{tabular}
 \label{table_thickness}
\end{table}

\begin{table}[h]
 \caption{Error in the inner pressure for different dilation levels with tube thickness $T$ = 0.2. Higher dilation level is simulated by adding more fluid in the tube, as shown by the increase of $U(r_\text{i})$. The relative error $\epsilon_p = \frac{|P_{\text{numerical}} - P_\text{analytic}|}{P_\text{analytic}}$}
 \centering
\begin{tabular}{l | l | l | l | l | l | l}
 \hline \hline
 Dilation level & $U(r_\text{i})$ & $P_{\text{numerical}}$ & $ P_\text{analytic}$ & $\epsilon_p$\\ [1ex]
 \hline
 Level 1 & 0.02827 & 159.7 & 161.24 & 9.57e-3\\
 Level 2 & 0.07989 & 399.4 & 403.14 & 9.28e-3\\
 Level 3 & 0.15658 & 652.7 &663.85 & 1.68e-2\\
 Level 4 & 0.23483 & 805.2 & 840.9 & 4.25e-2\\
 \hline
 \end{tabular}
 \label{table_level}
\end{table}

As can be seen from Table~\ref{table_thickness} and Table~\ref{table_level}, our fiber-based tube model is able to capture the analytical trend for various tube thickness and dilation levels, with relative error below 5\%.

\section{Esophageal transport}

In the previous section, we showed that our IB formulation for an elastic tube (with an appropriate arrangement of fibers) is able to capture the analytical trend of a dilation process. In this section, we extend the elastic tube model to describe esophageal transport. 

\begin{figure}
 \centering
 \includegraphics[scale = 0.4]{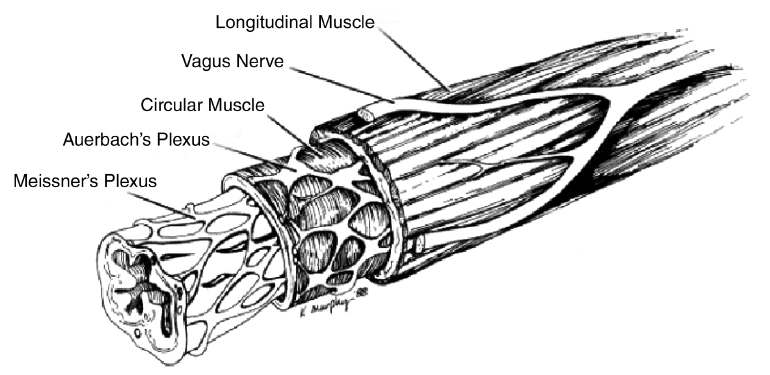} 
 \caption{Schematic of the multiple layers of the esophagus. The inner most layer is the mucosal layer (including mucosa and submucosa), which is highly folded at rest; the outer layers are muscle layers including circular muscle layer and longitudinal muscle layer (reproduced with permission from Kahrilas~\cite{Kahrilas1989}).}
 \label{fig-peter_image}
 \end{figure}

The human esophagus is a long multi-layered composite tube that consists of inner mucosal-submucosal 
layers (collectively referred to as ``mucosal" layer) and  outer muscle layers which in turn include 
the circular and longitudinal muscle layers (so named because of their fiber orientations~\cite{Kahrilas1989}).
The anatomy of the human esophagus is illustrated in Fig.~\ref{fig-peter_image}. In-vitro tests show that there exists a weak connecting tissue, referred to as the ``interfacial" layer, between the muscle and mucosal layers~\cite{Greg2000}. 
The bolus (depending on its content) is generally considered as a Newtonian fluid, with its viscosity varying 
from one centipoise (cP) to several hundred centipoise~\cite{Dooley1988}. The overall volume of the bolus is on the order of a few milliliters from the clinical study~\cite{dantas1990effect}. Studies of 
Pouderoux et al.~\cite{Kahrilas1997} and Mittal et al.~\cite{Mittal2006} suggest that 
during bolus transport, the circular muscle contraction is well coordinated 
with the longitudinal muscle shortening. It is this key activation pattern that is used 
in our esophageal muscle model that enables the transport of the ``tear'' shaped bolus~\cite{Li-Brasseur1993,Li1994} through the esophagus.

\subsection{Geometry, boundary conditions and material properties}

\begin{figure}
 \centering
 \includegraphics[scale = 0.4]{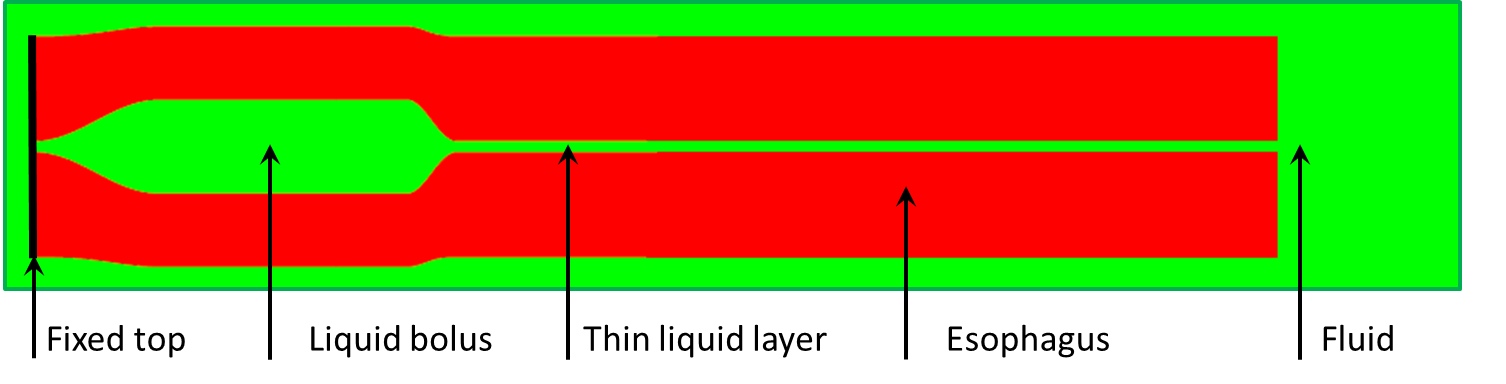} 
 \caption{Schematic (not drawn to scale) of the computational domain consisting of the elastic esophagus (red) and a viscous fluid (green). The elastic esophagus, a cylindrical tube with its top end fixed, is immersed in the background fluid in our 3D computational model. The upper esophagus is initially filled with a bolus, and the lower part is filled with a thin 
  liquid layer in the lumen. Traction-free boundary conditions are applied to all surfaces of the rectangular computational domain.}
 \label{fig_overall_scheme}
\end{figure}

The reference configuration of the esophagus model is taken to be a long straight cylindrical tube made up of elastic fibers. 
There are five important components in our esophagus model: (1) inner mucosal (IM) layer; (2) outer mucosal (OM) layer; (3) interfacial (IF)
layer; (4) circular muscle (CM); and (5) longitudinal muscle (LM). The IM and OM layers together 
represent the mucosal layer of the esophagus, which we split into two layers for numerical purposes (see Sec. \ref{sec_num_issues}). The length of the esophagus tube is taken to be 240 mm, 
as the typical human esophagus length is in the range of 180-250 mm~\cite{Meyer1986}. The thin liquid layer confined in the narrow esophageal lumen is assumed to have a circular cross section in the reference configuration with a radius of 0.3 mm. The thickness of each esophageal wall component is obtained based on the clinical data of human esophagus at non-rest state (i.e., with intruded catheter in the esophagus). The thickness of each layer at rest is listed in Table~\ref{table_esophagus_thickness}, based on 
the clinical data of Mittal et al.~\cite{Mittal2005}. The entire esophagus is immersed in a fluid region of size~$(-7~\text{mm},7~\text{mm}) \times (-7~\text{mm},7~\text{mm}) \times 
 (-25~\text{mm}, 245~\text{mm})$. On the six surfaces of the fluid box, we impose stress-free boundary conditions. We also fix the esophageal top end, which, in physiological situation, is constrained by the upper esophageal sphincter. The overall schematic is shown in Fig.~\ref{fig_overall_scheme}. We here consider the transport of an initially filled bolus in the upper end of esophagus. For all the cases presented here, we take uniform viscosity of 10 cP and uniform density of $\text{1g/cm}^3$ for both the fluid and the esophagus tube.

The esophageal tissue is generally modeled as a nonlinear anisotropic elastic or pseudo-elastic material.
The reported material properties (such as the modulus of the muscle layers in the circumferential and longitudinal orientations) are, however, substantially different~\cite{Fan2004,Natali2009,Sokolis2013,Yang2006a,Yang2006b,Stavropoulou2009}. 
Here we assume an elastic behavior of each fibrous esophageal layer. Such models have been used before to estimate the muscle tension~\cite{Nicosia-Brasseur2002}. 
The elastic modulus for the radial, circumferential, and axial fibers is taken as $S = 4$~kPa. 
The interfacial layer that loosely connects the muscle and mucosal layer consists of only radial fibers 
with modulus $S=0.0004$~kPa. For the mucosal layer, Stavropoulou~\cite{Stavropoulou2009} characterized the elasticity of the mucosal layer in their axial and circumferential directions, which can be represented by 
axially-circumferentially-radially arranged fiber network. By contrast, Natali et al.~\cite{Natali2009} considered the
mucosal layer to comprise two families of helical fibers embedded in a matrix tissue. Thus, to demonstrate the capabilities of our modeling approach, we consider both fiber arrangements of the mucosal 
layer in Sections~\ref{sec_case_axial_mucosal} and~\ref{sec_case_helical_mucosal}.

\begin{table}[ht]
 \caption{Thickness of each esophageal layer: clinical data~\cite{Mittal2005} and data used in our model. Note that the clinical test obtains the thickness of each layer at its non-rest state, with esophageal lumen dilated by the intruded catheter. Computer model adopts the thickness of each layer at its rest state. The thickness of mucosal layer measured based on the ultrasound image is much lower than the thickness at rest, as the intruded catheter will distend the mucosal layer significantly and reduce the layer thickness.}
 \centering
\begin{tabular}{l | l | l | l | l | l}
 \hline \hline
 Unit (mm) & \specialcell{Lumen \\ radius} & \specialcell {Mucosa \\ thickness} & \specialcell {IF \\ thickness} & \specialcell {CM \\ thickness} & \specialcell {LM \\ thickness} \\ 
 \hline 
 Clinical data & 3.5 & 1.85 & NA &0.55 &0.49 \\
 Computer model & 0.3 & 3.2 & 0.6 & 0.6 & 0.6 \\ 
 \hline
 \end{tabular}
 \label{table_esophagus_thickness}
\end{table}

\subsection{Muscle activation}
\label{sec_muscle_active}
In-vivo experiments~\cite{Kahrilas1997,Mittal2006} show that during normal esophageal 
transport, there is well-coordinated circular muscle (CM) contraction and longitudinal muscle (LM) shortening. 
A quantitative model to characterize the contraction and shortening process in terms of neuronal firing
or reaction kinetics in muscles is not available; however, experiments show that there is a precise 
synchrony between the two types of muscle activation patterns~\cite{Kahrilas1997,Mittal2006}. 
In our model, this sequential activation is implemented by dynamically changing the rest lengths of 
springs. Specifically, let $z$ denote the vertical coordinate based on the initial configuration
of the esophageal tube, with the bottom end of the esophagus as the origin $z=0$, and the top as the end $z=L$. Then an active spring representing a section of one active muscle fiber 
has its rest length $r(z,t)$ given by
\begin{equation}
r(z,t) = \begin{cases} r_0 & \mbox{if } t-t_0 \leq \frac{L-z}{c} \\ 
(1-a(z,t))r_0 & \mbox{if } \frac{L-z}{c} < t-t_0 < \frac{L-z}{c} + \frac{\Delta L}{c} \\
r_0 & \mbox{if }  t- t_0 \geq \frac{L-z}{c} + \frac{\Delta L}{c} \end{cases} \label{eqn_contract_spring} 
\end{equation}

in which, $r_0$ is the spring's initial rest length, $c$ is the speed of the activation wave, $t_0$ is the 
initiation time of activation, $a(z,t)$ is the reduction ratio, and $\Delta L$ is the contracting segment's length in the reference coordinate system. Eq.~\eqref{eqn_contract_spring} gives the rest length of a spring at its rest, activation and relaxation state, respectively. The equation also shows, at any time, the whole esophageal tube has a contracting segment with a vertical length $\Delta L$. The variation of muscle activation along this contracting segment will likely influence bolus transport. To understand this influence, we propose two muscle activation models, namely \textit{uniform muscle activation} and \textit{nonuniform muscle activation}. They differ by how the reduction ratio $a(z,t)$ is distributed along the contracting segment $\Delta L$:

\textit{Uniform muscle activation}: 
\begin{equation}
a(z,t) = a_0 \label{eqn_uniform_contraction}
\end{equation}
\textit{Nonuniform muscle activation}: 
\begin{equation}
a(z,t) = a_0 e^{-0.5(z-z_0(t))^2/{\lambda}^2}, \label{eqn_nonuniform_contraction}
\end{equation}
where $a_0$ is a constant, $z_0(t)$ is the $z$-coordinate at the vertical center of the contraction segment, and $\lambda$ is the parameter that controls the width of the Gaussian distribution in eq.~\eqref{eqn_nonuniform_contraction}.  The common parameters of muscle activation model used in all the cases of esophageal transport are listed in Table~\ref{tab_activation_para}. $a(z,t)$ differs in test cases depending on whether uniform muscle activation or nonuniform muscle activation is used.
\begin{table}[ht]
 \caption{Model parameters for the circular muscle (CM) contraction and longitudinal muscle (LM) shortening used in all the cases. The muscle 
 activation model is based on eq.~\eqref{eqn_contract_spring}. }
 \centering
\begin{tabular}{l | l | l | l | l }
 \hline \hline
 Muscle activation type  & $c$ (mm/s)  & $\Delta L$ (mm) & $t_0$ (s) \\ [1ex]
 \hline
 CM contraction  & 100 & 60 & 0 \\
 LM shortening  & 100 & 60 & 0 \\ 
 \hline
 \end{tabular}
 \label{tab_activation_para}
\end{table}

\subsection{Numerical issues}
\label{sec_num_issues}
Esophageal transport involves multiple length scales, which is evidenced by the fact that 
the esophageal length is 240 mm, while the lumen radius at rest is only 0.3 mm. The requirement of resolving 
the narrow lumen dictates the grid size of the problem. More challenging in terms of computational 
modeling is the large deformation of the inner lumen, due to the dilation caused by bolus movement. This can be 
seen in Fig.~\ref{fig-CSA} (Section~\ref{sec_case_axial_mucosal}). For good numerical resolution of the 
transport problem, we use $\Delta x = \Delta y=0.2 ~\text{mm and } \Delta z =1 ~\text{mm}$ for the Eulerian 
mesh. The choice of the Lagrangian mesh size must also address the issues of large deformations of lumen and 
the external fluid leaking into the tube, which could occur when the Lagrangian mesh becomes coarser than its 
Eulerian counterpart~\cite{CSPeskin02}. In our implementation, considering the fact that significant deformation 
is mainly confined in the inner mucosa (IM) layer, we employ a refined mesh in the IM layer along the 
circumferential and axial orientations to form an ``impermeable" surface. For the outer layers with relatively 
small dilation, a relatively coarser Lagrangian mesh is used to reduce the computational cost. The mesh sizes 
for various esophageal components are listed in Table~\ref{tab_solid_mesh_size}. The axial beams included 
in the model provide resistance to curvature changes in the axial direction that are associated with buckling of the tube. The 
time step $\Delta t$ needs to satisfy the stability constraints from both the fluid and solid system. Based on 
empirical tests, we choose $\Delta t = 0.02$~ms. The total time for the transport is about 2.5 s, which requires 
about 125,000 timesteps. The relative change in the bolus volume is within 0.8\% before the bolus begins to 
empty from the bottom of the esophagus. This indicates that the immersed esophagus model is relatively ``water-tight''.

\begin{table}[ht]
 \caption{Grid size along $r, \theta, z$ orientations (denoted as $\Delta_r, r\Delta_\theta, \Delta_z$) for each layer of the esophagus in the reference configuration. Note that the grid size in circumferential orientation $r\Delta_\theta$ for each layer should be understood as an average value, since the coordinate $r$ increases from the inner surface to the outer surface while $\Delta_\theta$ is constant in each layer.}
 \centering
\begin{tabular}{l | l | l | l | l | l}
 \hline \hline
 Grid size (0.1mm) & IM & OM & IF & CM & LM \\ [1ex]
 \hline
  $\Delta_r$& 2 & 4 & 3 & 3 & 2 \\
 $r\Delta_\theta$ & 0.5 & 2 & 1.9 & 2.2 & 2.5 \\ 
	$\Delta_z$ & 4 & 8 & 8 & 8 & 8 \\ 
 \hline
 \end{tabular}
 \label{tab_solid_mesh_size}
\end{table}

\section{Results}

\subsection{Case 1: Axially-circumferentially-radially arranged mucosal fibers with uniform muscle activation} % 07/07/2014 modify by walter
\label{sec_case_axial_mucosal}
Here we report a case study of the esophageal transport that considers the mucosal layer to be composed 
of axially-circumferentially-radially arranged fibers. Experiments show the intact mucosal layer is highly
folded at rest (see Fig. 1 in Ref.~\cite{Yang2006a}), so we take a relatively low moduli for circumferential,
radial, and axial fibers as 0.004~kPa, 0.004~kPa and 0.04~kPa, respectively, to qualitatively capture its 
effective elastic property. We use uniform activation model as described in eq.~\eqref{eqn_uniform_contraction}. The reduction ratio $a_0$ for CM contraction and LM shortening is taken to be
0.6 and 0.5, respectively. Other parameters are listed in Table~\ref{tab_activation_para}. 

\begin{figure}[ht]
 \centering
 \includegraphics[scale = 0.35]{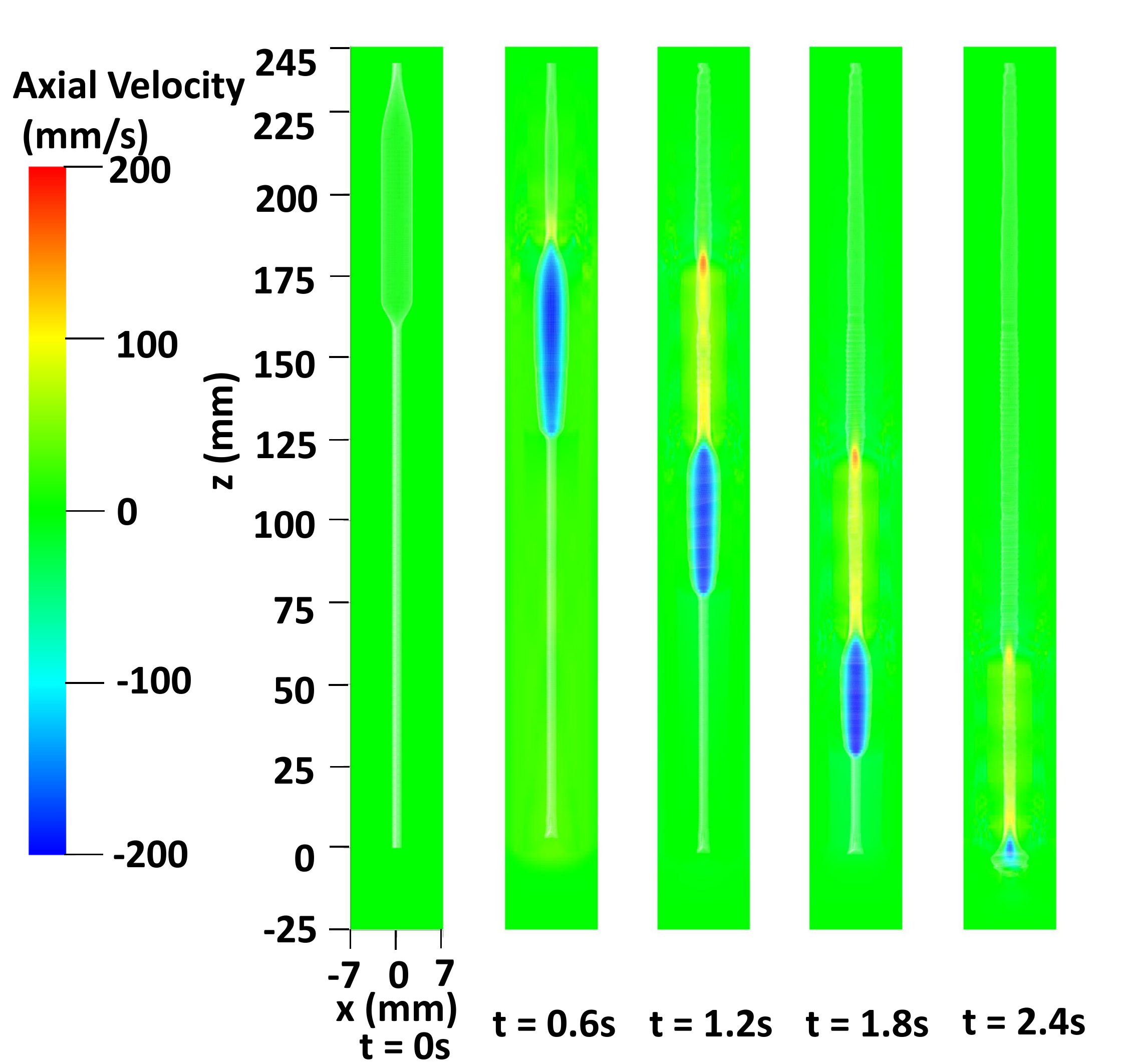}
 \caption{Axial velocity in the plane $y=0$ at different times for Case 1 in Section~\ref{sec_case_axial_mucosal}. Only the inner mucosal (IM) layer (white) of the esophagus is shown to better visualize the inside bolus.}
 \label{fig-uz}
\end{figure}

\begin{figure}[ht]
 \centering
 \includegraphics[scale = 0.35]{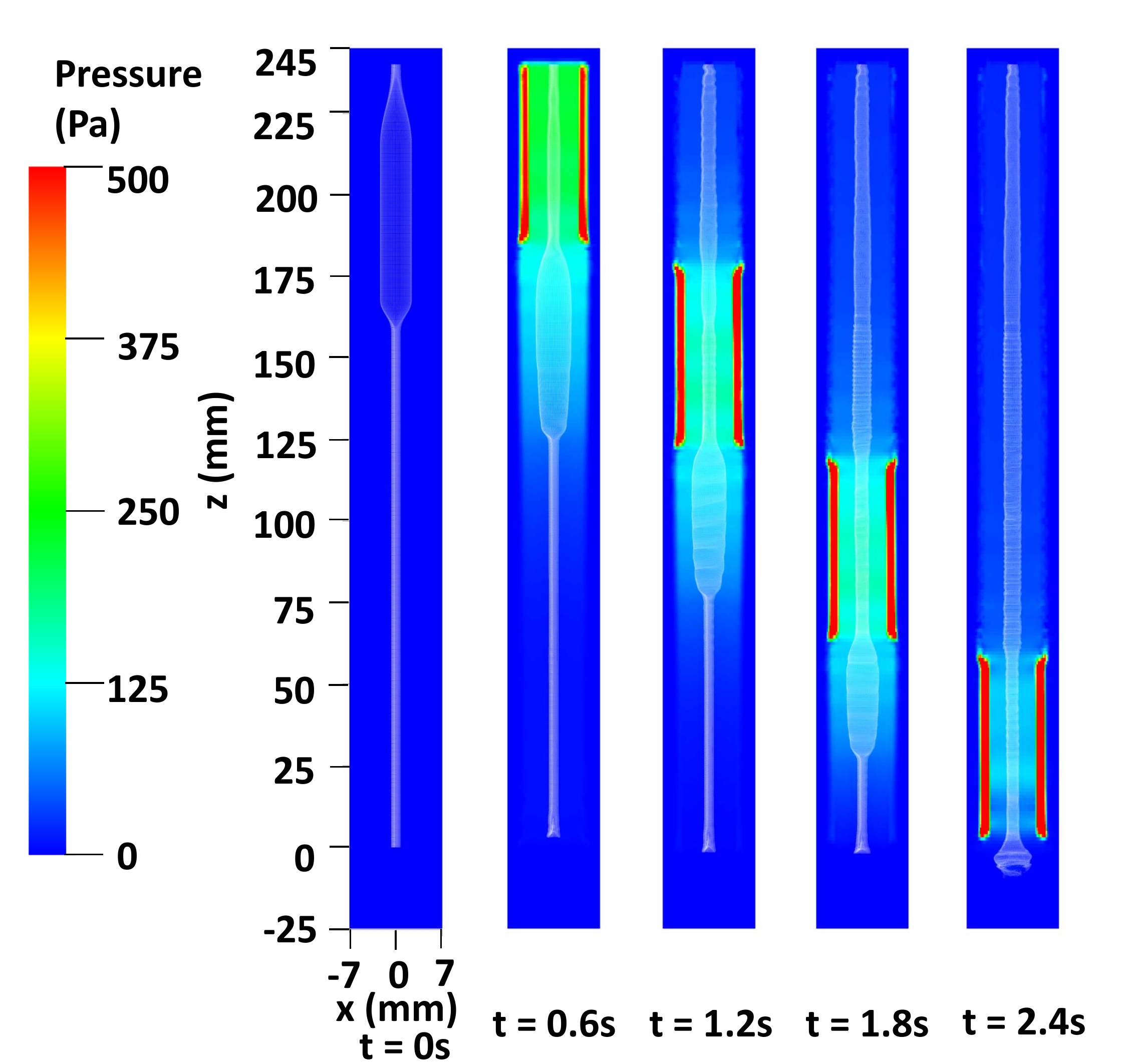}
 \caption{Pressure in the plane $y=0$ at different times for Case 1 in Section~\ref{sec_case_axial_mucosal}. Only the inner mucosal (IM) layer (white) of the esophagus is shown to better visualize the inside bolus.}
 \label{fig-pressure}
\end{figure}

As shown in the Fig.~\ref{fig-uz}, the bolus is transported to and emptied through the bottom of the esophagus 
in about 2.4 seconds. The running bolus (indicated by the negative axial velocity) is confined by the inner 
mucosal layer. This underlines the role of mucosal layers in preparing the ``tear-drop" shape of the bolus. 
The pressure distribution shown in Fig.~\ref{fig-pressure} implies that the primary pumping force behind 
the bolus is generated by muscle activation, evidenced by the peak pressure at the contraction region. 
This is consistent with reported experimental observations~\cite{Mittal2006}. We remark that  
the ``tear-drop" shape of the running bolus is not specified, but rather is a consequence of the 
fluid-structure-muscle activation interaction. This is different from previous models for bolus 
transport~\cite{Li-Brasseur1993,Ghosh2005,Nicosia-Brasseur2002} where the bolus shape was pre-defined. 

\begin{figure}[ht]
 \centering
 \includegraphics[scale = 0.25]{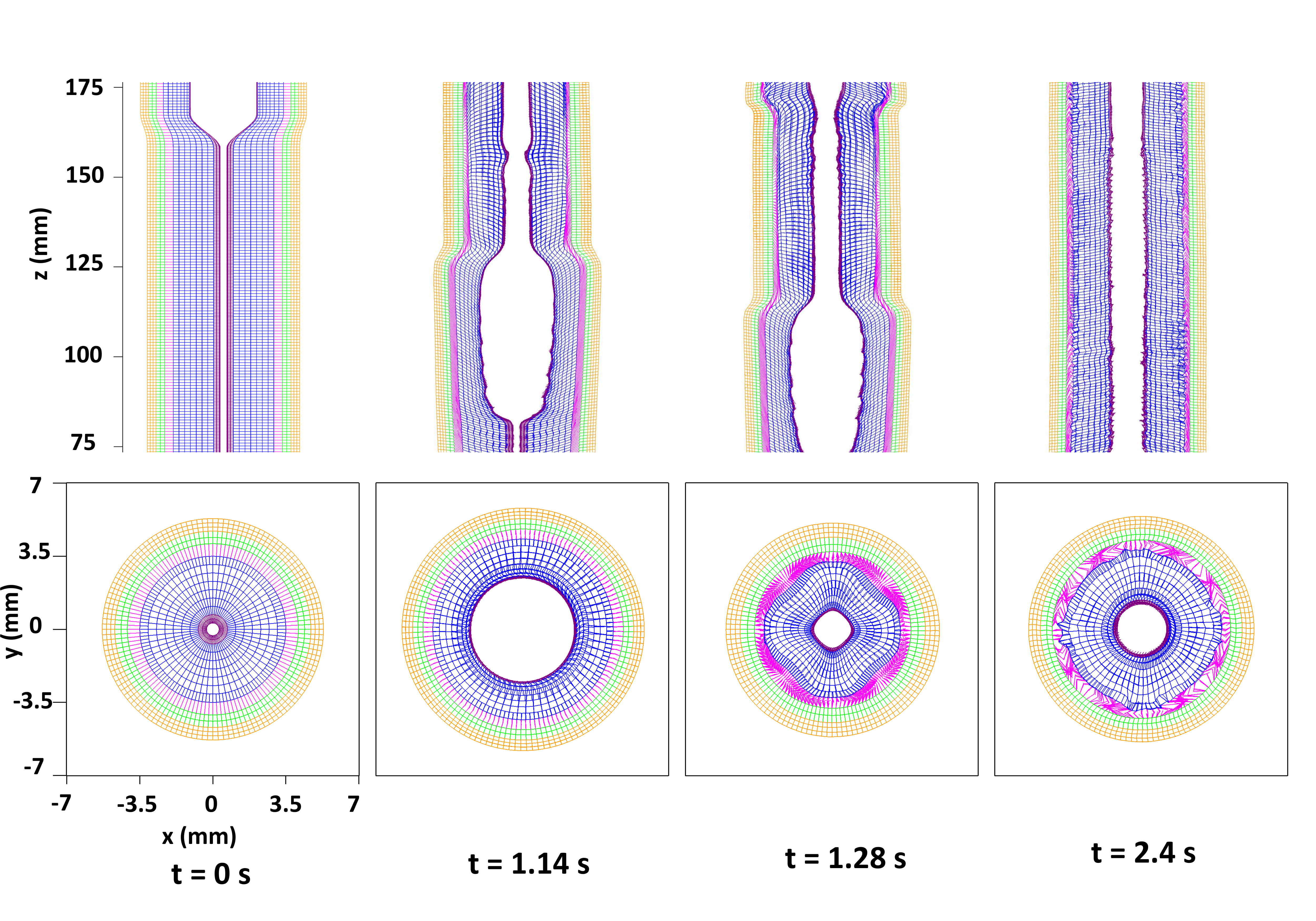}
 \caption{Kinematics of esophageal layers at four different stages: at rest ($t$=0 s); at dilation 
 ($t$=1.14 s); at contraction ($t$=1.28 s); and at relaxation ($t$=2.4 s) for Case 1 in Section~\ref{sec_case_axial_mucosal}. Purple, blue, magenta, green and orange meshes from the inside to the outside, denote the 
 IM, OM, IF, CM and LM layers, respectively. (Upper) Side view of a section of the esophagus within
 the box: $(-7~\text{mm},7~\text{mm}) \times (-0.2~\text{mm},0.2~\text{mm}) \times 
 (75~\text{mm}, 175~\text{mm})$; (Lower) top view of a section of the esophagus within the box:
 $(-7~\text{mm},7~\text{mm}) \times (-7 ~\text{mm},7 ~ \text{mm})
 \times (119.5 ~\text{mm}, 120.5 ~\text{mm})$. Because of complex kinematics of the esophageal structure, the apparently overlapping or missing springs in the above figures are actually a consequence of the out-of-plane motions of the springs. 
}
 \label{fig-CSA}
\end{figure}

\begin{figure}[ht]
 \centering
 \includegraphics[scale = 0.35]{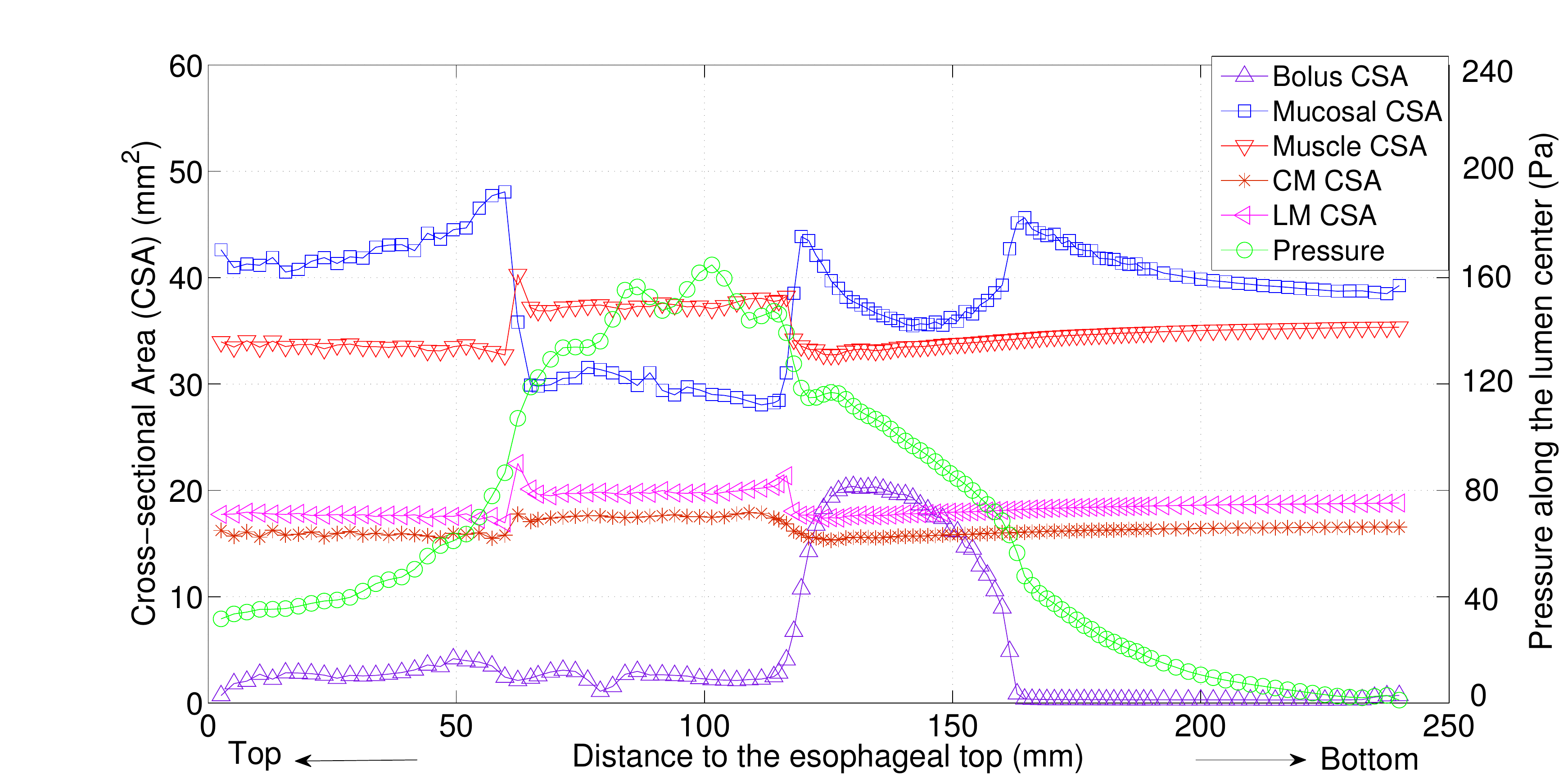}
 \caption{The cross-sectional area (CSA) of the bolus and the esophageal components, and the lumen 
 pressure along its central line: $x=0,y=0$, at $t=1.2$ s for Case 1 in Section~\ref{sec_case_axial_mucosal}. }
 \label{fig-CSA_pressure}
\end{figure}

Detailed information on the deformation of each esophageal layer is illustrated in Fig.~\ref{fig-CSA}. It can 
be seen that a typical esophageal segment (at each axial location) passes through four distinct stages: (a) the segment is at rest; (b) the segment is dilated by the incoming bolus; (c) the segment contracts as a result of the incoming activation wave; and (d) the segment relaxes after the activation wave passes. Figs.~\ref{fig-uz} and~\ref{fig-CSA} show that the mucosal layer plays an important role in the bolus transport. 
First, the inner layer of mucosa shapes the running bolus by dynamically closing (or narrowing) the lumen 
above the bolus region while opening the lumen from below. Second, the pronounced axial movement of 
mucosal layer ``lubricates" the running bolus, which helps to achieve  better transport 
efficiency. Previous studies on bolus transport excluded mucosal layers, and did not capture this important feature of bolus movement~\cite{Li-Brasseur1993,Ghosh2005,Nicosia-Brasseur2002}. Quantitative results for the cross-sectional area (CSA) of the esophageal layers, the bolus, as well as for the
pressure distribution along the lumen center at time instant $t=1.2$ s are shown in Fig.~\ref{fig-CSA_pressure}. 
At the contraction region behind the bolus, a pressure peak and an increased muscle CSA (i.e. the sum of CSA of LM layer and CSA of CM layer) coexist, which is consistent
with the experimental observation of Mittal et al.~\cite{Mittal2006}.

% --------------------------------------------------- add new section: nonuniform contraction wave
\subsection{Case 2: Axially-circumferentially-radially arranged mucosal fibers with non-uniform muscle activation} % 07/07/2014 modify by walter
\label{sec_case_axial_mucosal_nonuniform}

\begin{figure}[ht]
 \centering
 \includegraphics[scale = 0.35]{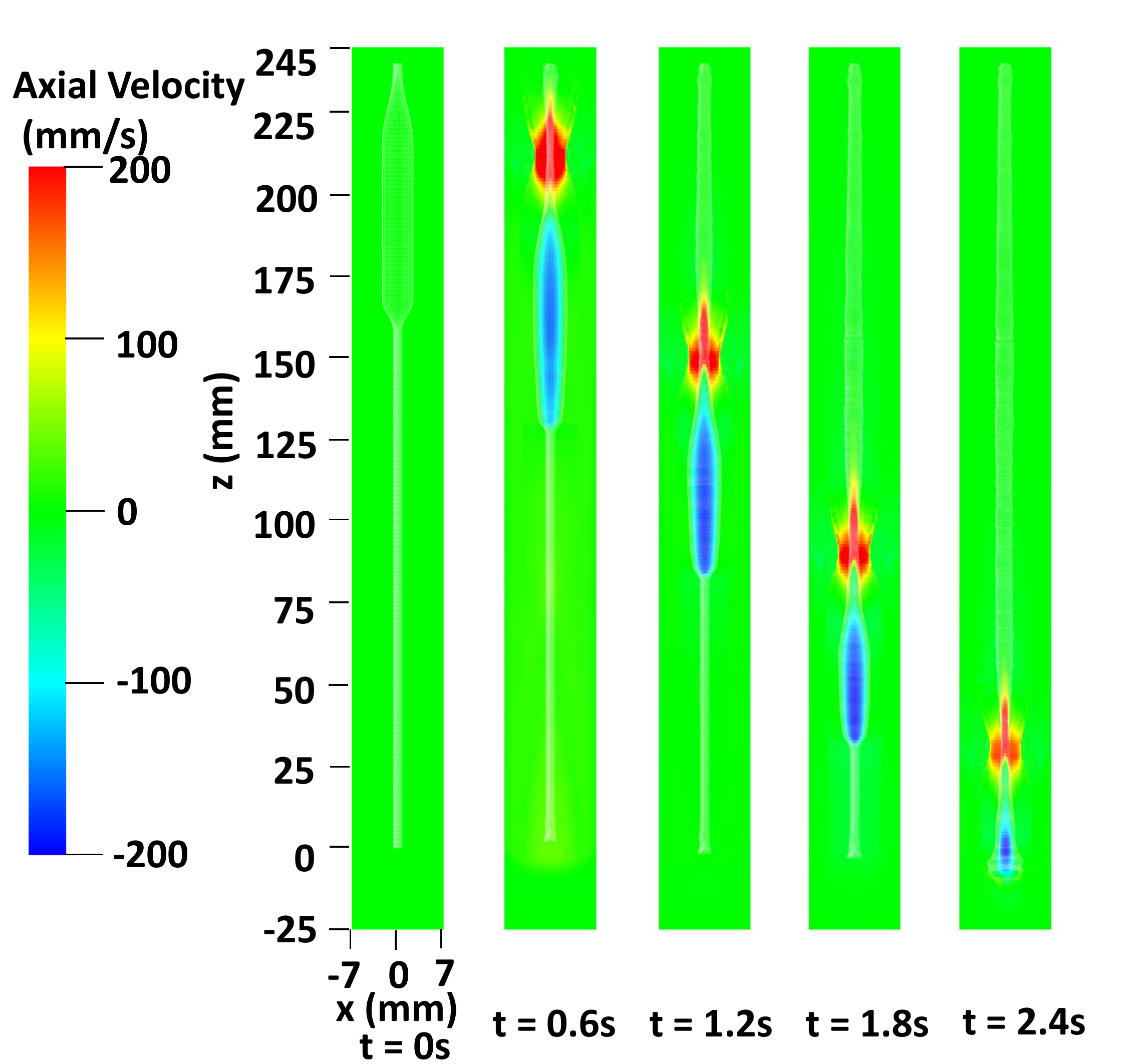}
 \caption{Axial velocity in the plane $y=0$ at different times for the Case 2 in Section~\ref{sec_case_axial_mucosal_nonuniform}. Only the inner mucosal (IM) layer (white) of the esophagus is shown to better visualize the inside bolus.}
 \label{fig-uz_nonuniform}
\end{figure}

\begin{figure}[ht]
 \centering
 \includegraphics[scale = 0.35]{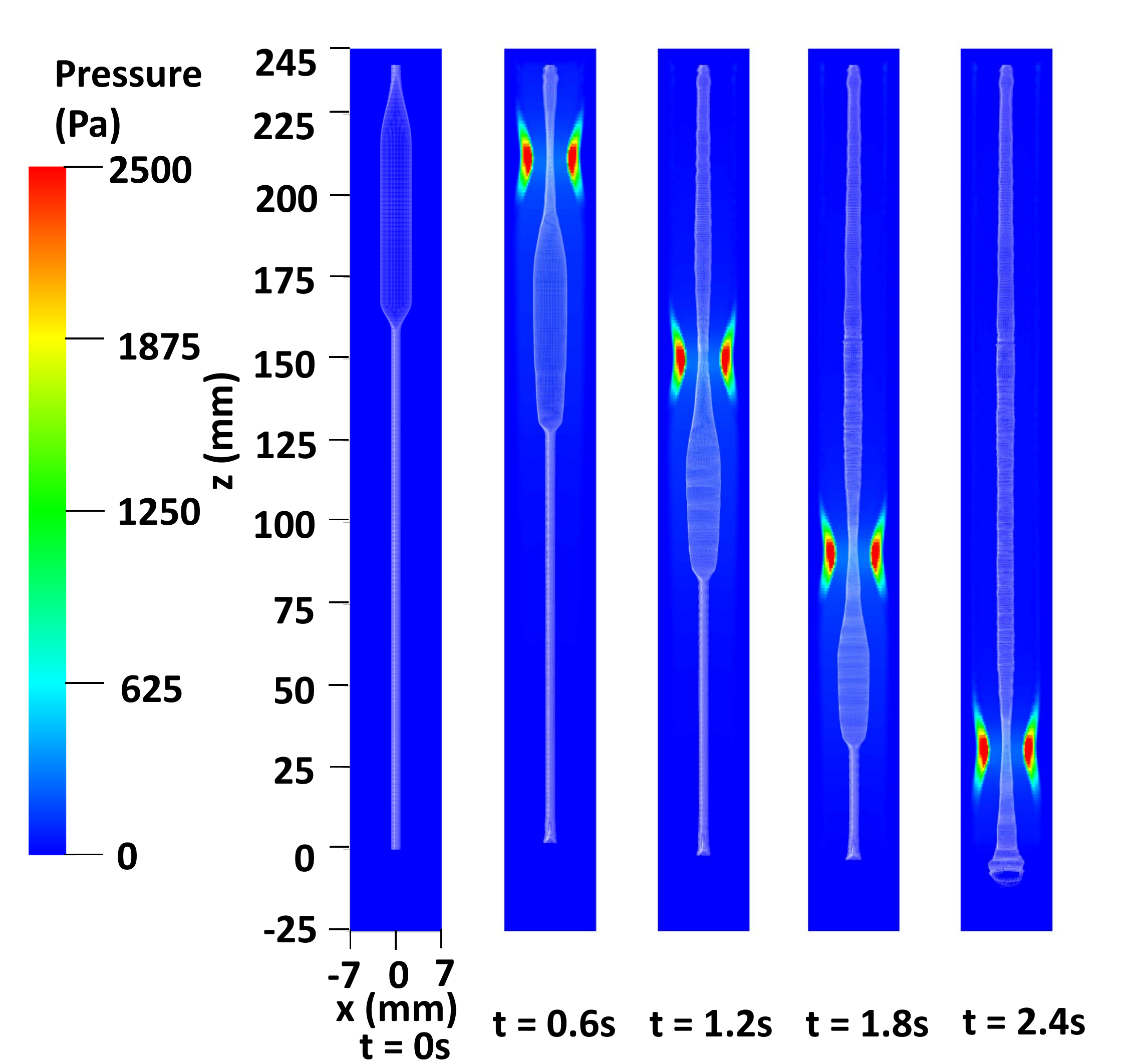}
 \caption{Pressure in the plane $y=0$ at different times for the Case 2 in Section~\ref{sec_case_axial_mucosal_nonuniform}. Only the inner mucosal (IM) layer (white) of the esophagus is shown to better visualize the inside bolus.}
 \label{fig-pressure_nonuniform}
\end{figure}

\begin{figure}[ht]
 \centering
 \includegraphics[scale = 0.25]{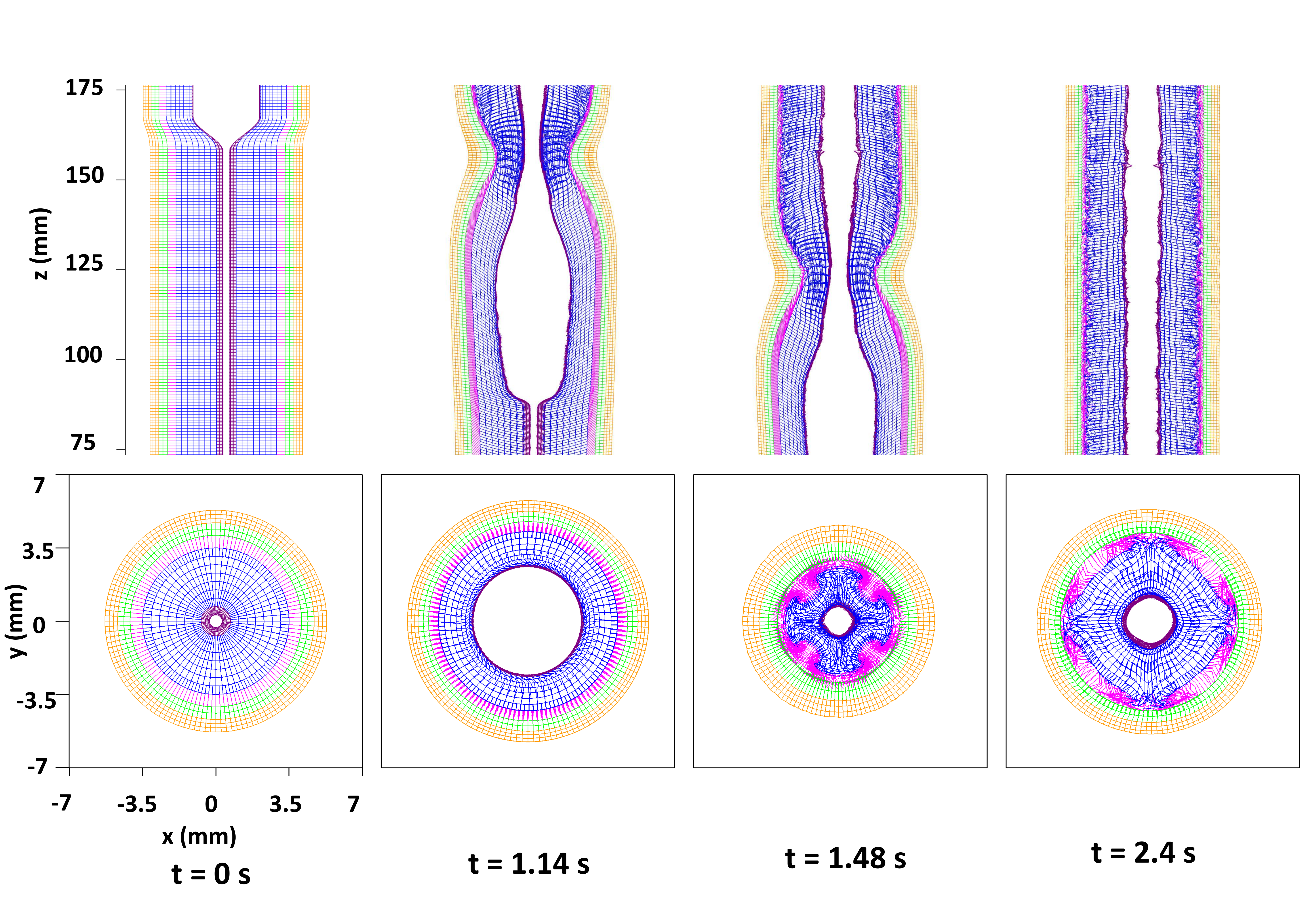}
 \caption{Kinematic information of esophageal layers at four different stages: at rest ($t$=0 s); at dilation 
 ($t$=1.14 s); at contraction ($t$=1.48 s); at relaxation ($t$=2.4 s), for the Case 2 in Section~\ref{sec_case_axial_mucosal_nonuniform}. Purple, blue, magenta, green and orange meshes from the inside to the outside, denote the 
 IM, OM, IF, CM and LM layers, respectively. (Upper) Side view of a section of the esophagus within
 the box: $(-7~\text{mm},7~\text{mm}) \times (-0.2~\text{mm},0.2~\text{mm}) \times 
 (75~\text{mm}, 175~\text{mm})$; (Lower) top view of a section of the esophagus within the box:
 $(-7~\text{mm},7~\text{mm}) \times (-7 ~\text{mm},7 ~ \text{mm})
 \times (119.5 ~\text{mm}, 120.5 ~\text{mm})$. }
 \label{fig-CSA_nonuniform}
\end{figure}

\begin{figure}[ht]
 \centering
 \includegraphics[scale = 0.35]{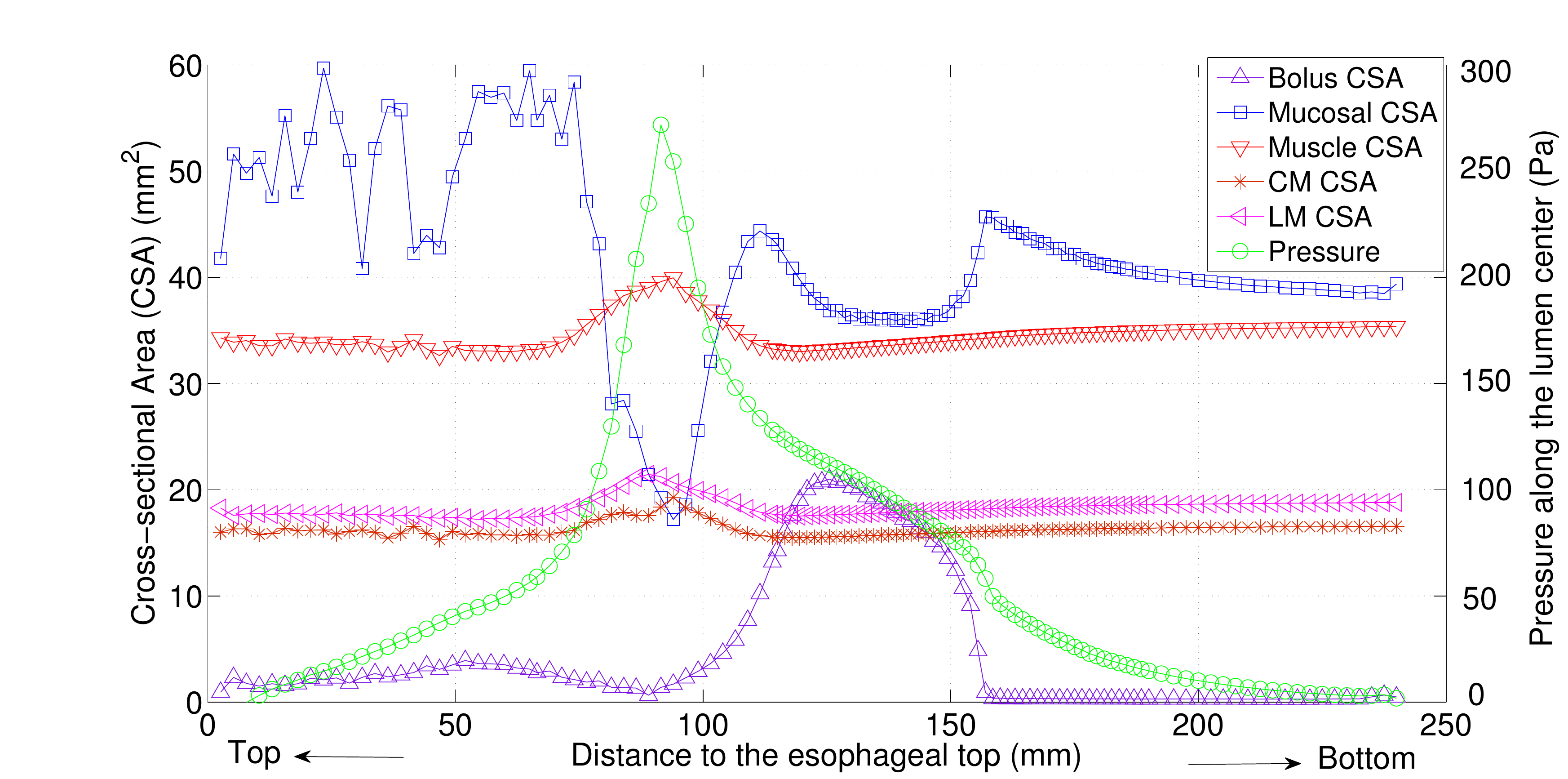}
 \caption{The cross-sectional area (CSA) of the bolus and the esophageal components, and the lumen 
 pressure along its central line: $x=0,y=0$, at $t=1.2$ s for Case 2 in Section~\ref{sec_case_axial_mucosal_nonuniform}. }
 \label{fig-CSA_pressure_nonuniform}
\end{figure}

Here we present the second case study of esophageal transport. We use the same geometry and material
model of esophagus as of Case 1 in Section ~\ref{sec_case_axial_mucosal}, except that we use nonuniform muscle activation model.  
We take $\lambda$ as 10 mm for both LM shortening and CM contraction, and we set reduction ratio $a_0$ for CM contraction 
and LM shortening as 0.6 and 0.5, respectively. Other parameters used in the muscle activation model 
are listed in Table~\ref{tab_activation_para}. The transport phenomenon is shown in Fig.~\ref{fig-uz_nonuniform} and Fig.~\ref{fig-pressure_nonuniform}. It is clear that the nonuniform muscle activation results in more pronounced contraction, as illustrated in Fig.~\ref{fig-CSA_nonuniform}. We remark that a distinctive muscle 
CSA peak overlaps with the pressure peak in this case, as shown in Fig.~\ref{fig-CSA_pressure_nonuniform}.
This is different from Case 1 in Section~\ref{sec_case_axial_mucosal}, which resulted in a plateau of increased muscle CSA 
(see Fig.~\ref{fig-CSA_pressure}).  A coexisting muscle CSA  and pressure peak is also observed in the 
clinical test of Mittal et al.~\cite{Mittal2006}, which, in conjunction with our simulation, implies that a 
synchronous nonuniform LM shortening and CM contraction probably corresponds to the normal physiology of esophageal 
muscle activation.

%---------------------------------------------------- add new section: end ---------------

\subsection{Case 3: Helical mucosal fibers with uniform muscle activation}
\label{sec_case_helical_mucosal}

\begin{figure}[ht]
 \centering
 \subfigure[]{
 \includegraphics[scale = 0.25]{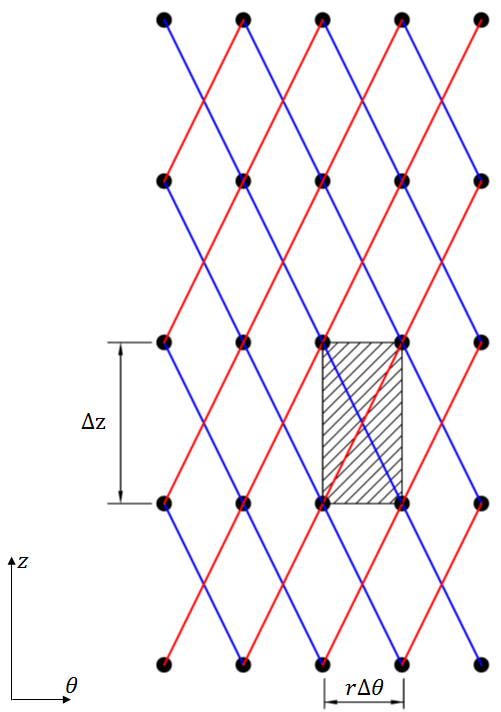}
 \label{fig-helix_rt}
 }
 \subfigure[]{
 \includegraphics[scale = 0.15 ]{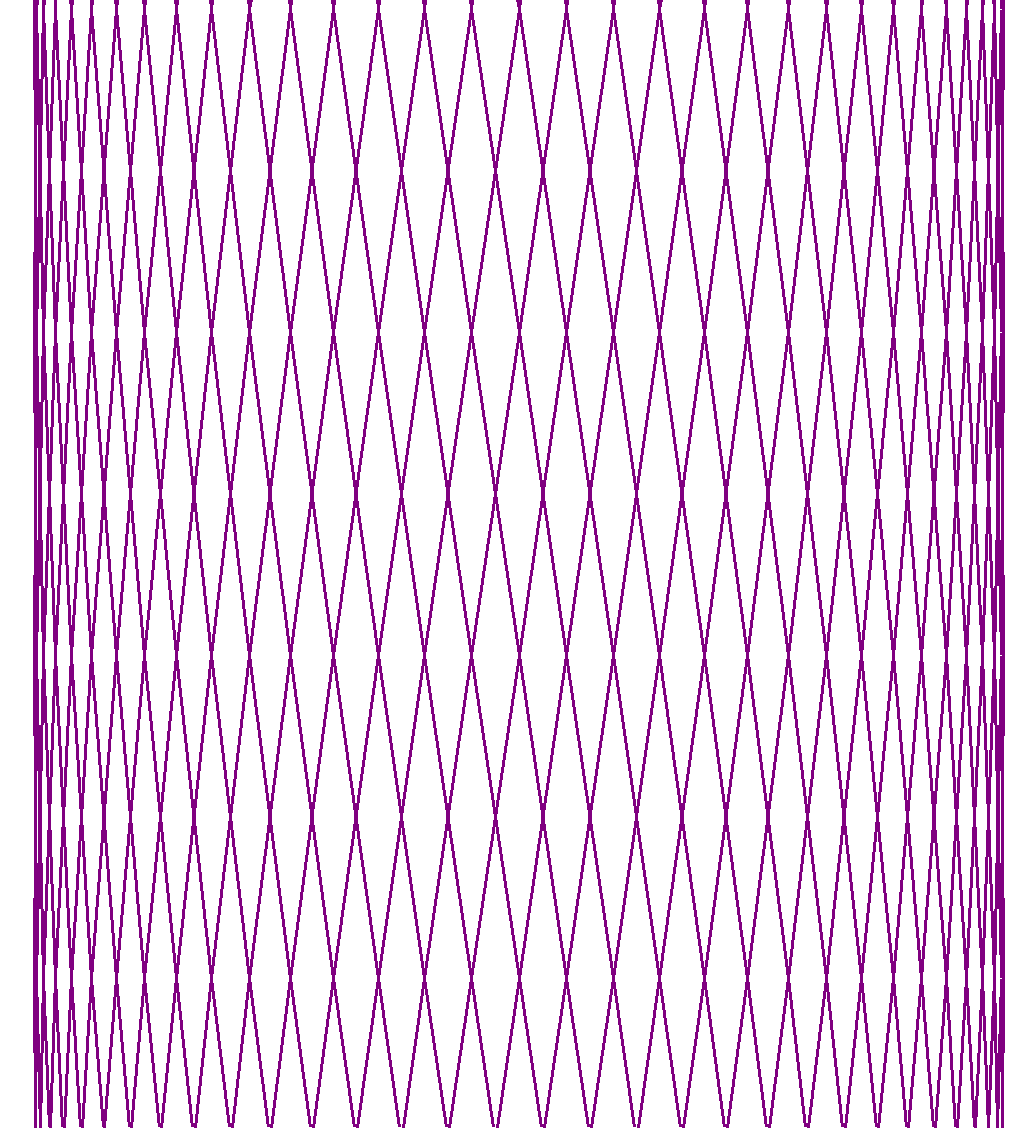}
 \label{fig-helix-IM}
 } 
 
 \caption{\subref{fig-helix_rt} Schematic of two families of helical fibers 
 (the blue and red lines, respectively) and the patch (the hatched area) in $(z,\theta)$ surface 
 with radial coordinate $r$. The patch volume associated with the helical spring of each family is 
 $r \Delta \theta \Delta r \Delta z$, if $a<r<b$; or $0.5r\Delta \theta \Delta r \Delta z$, if $r=a$ 
 or $r=b$, where $a$ and $b$ are the $r$-coordinate of the inner surface and outer surface, respectively. 
 \subref{fig-helix-IM} Illustration of helical fibers running on the inner-most surface of the mucosal layer.}
 \label{fig-helix}
\end{figure}

\begin{figure}[ht]
 \centering
 \includegraphics[scale = 0.35]{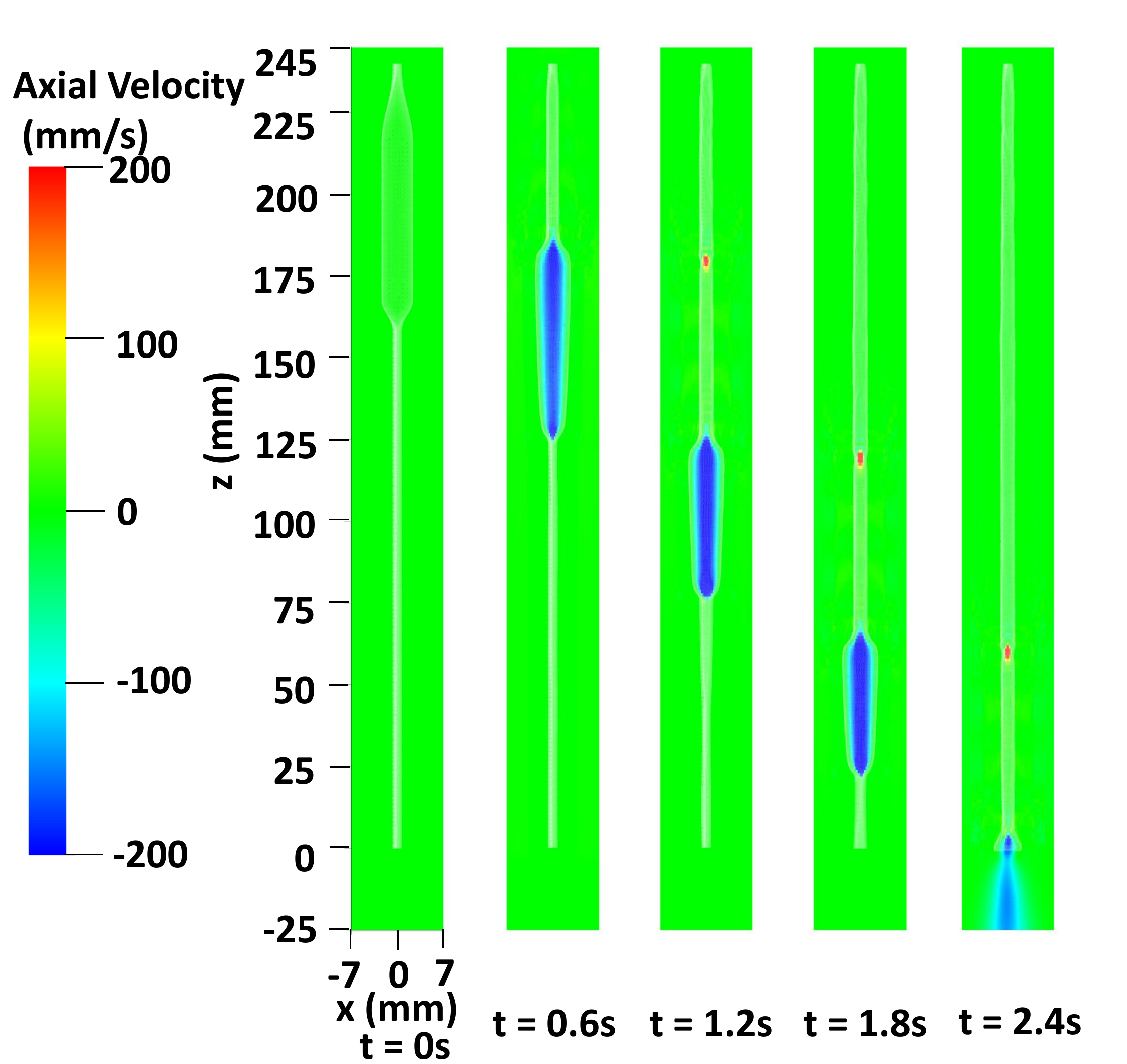}
 \caption{Axial velocity in the plane, $y=0$ at different times for the Case 3 in Section~\ref{sec_case_helical_mucosal}. Only the inner mucosal (IM) layer (white) of the esophagus is shown to better visualize the inside bolus.}
 \label{fig-helix-Uz}
\end{figure}

\begin{figure}[ht]
 \centering
 \includegraphics[scale = 0.35]{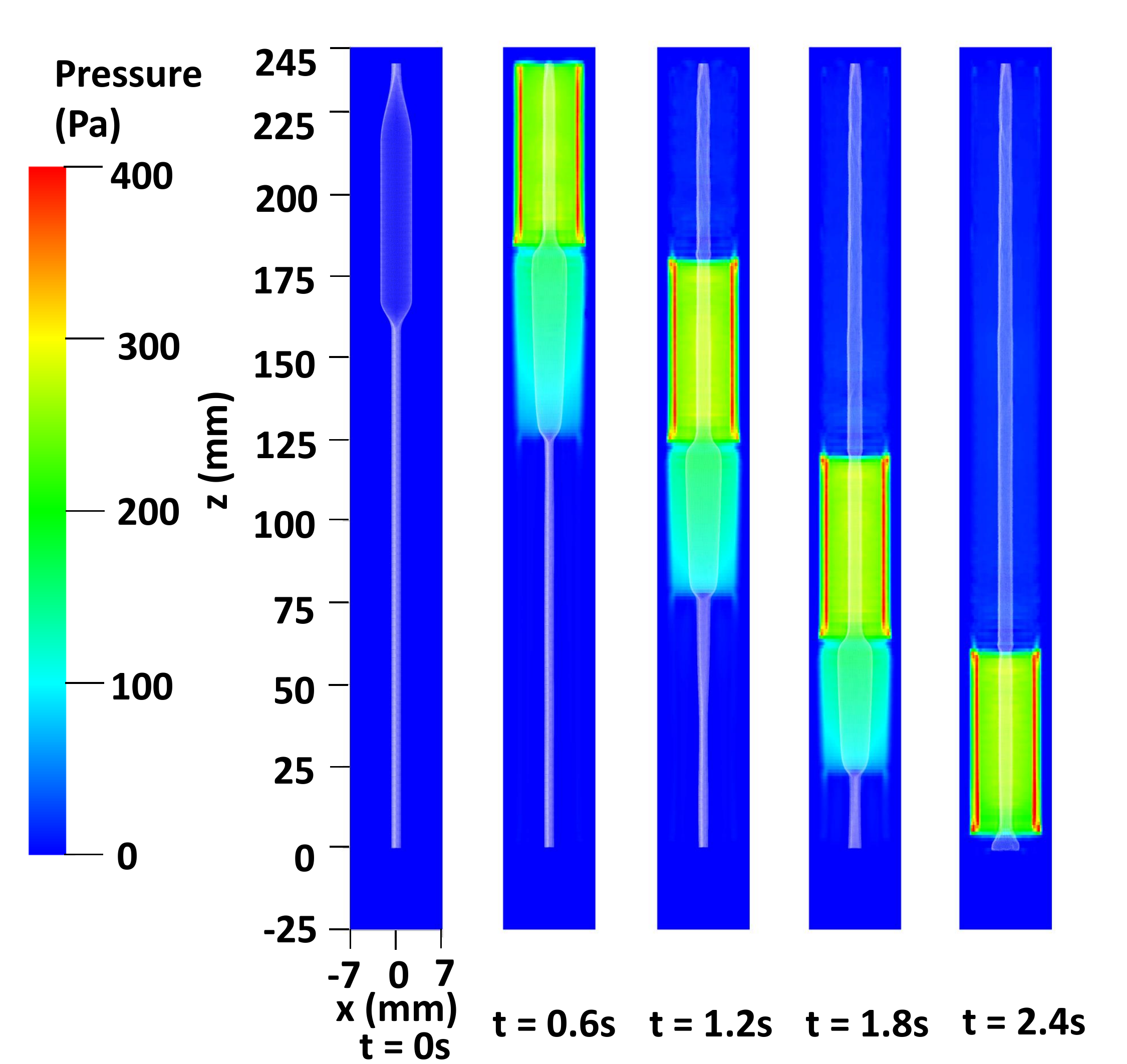}
 \caption{Pressure field in the plane, $y=0$ at different times for the Case 3 in Section~\ref{sec_case_helical_mucosal}. Only the inner mucosal (IM) layer (white) of the esophagus is shown to better visualize the inside bolus.}
 \label{fig-helix-pressure}
\end{figure}

\begin{figure}[ht]
 \centering
 \includegraphics[scale = 0.25]{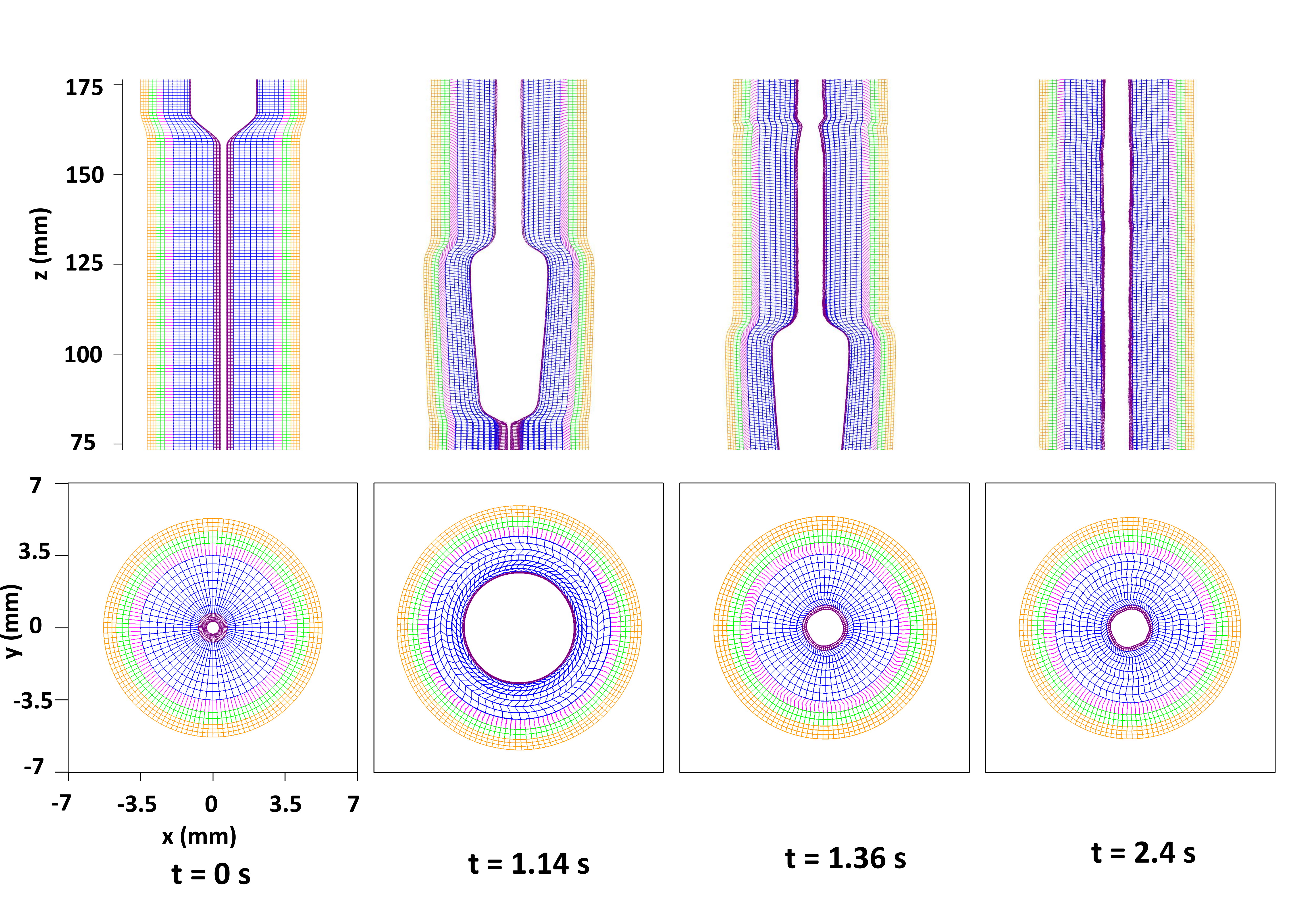}
 \caption{Kinematic information of esophageal layers at four different stages: at rest ($t$=0 s); at dilation 
 ($t$=1.14 s); at contraction ($t$=1.36 s); at relaxation ($t$=2.4 s), for the Case 3 in Section~\ref{sec_case_helical_mucosal}. Purple, blue, magenta, green and orange meshes from the inside to the outside, denote the 
 IM, OM, IF, CM and LM layers, respectively. (Upper) Side view of a section of the esophagus within
 the box: $(-7~\text{mm},7~\text{mm}) \times (-0.2~\text{mm},0.2~\text{mm}) \times 
 (75~\text{mm}, 175~\text{mm})$; (Lower) top view of a section of the esophagus within the box:
 $(-7~\text{mm},7~\text{mm}) \times (-7 ~\text{mm},7 ~ \text{mm})
 \times (119.5 ~\text{mm}, 120.5 ~\text{mm})$. }
 \label{fig-helix-csa}
\end{figure}

Natali et al.~\cite{Natali2009} proposed a constitutive model of a multi-layered esophagus, in 
which the mucosal layer consists of two families of helical fibers, and the muscle layer 
is composed of axially and circumferentially arranged fibers. Here we present a case with such a helical mucosal fiber arrangement. Two families of helical fibers run in the $(r,\theta)$ surfaces, as shown in 
Fig.~\ref{fig-helix}. Radial springs are used to link the $(r,\theta)$ surfaces. The 
modulus of the helical fibers is taken to be the same as that of fibers in the muscle layer, while the modulus of the radial 
fibers of the mucosal layer are one order of magnitude weaker than the modulus of helical fibers. We use uniform muscle activation model with reduction ratio $a_0$ for CM contraction and LM shortening as 0.5 and 0.3, respectively. Other parameters of muscle activation model are listed in Table~\ref{tab_activation_para}. Fig.~\ref{fig-helix-Uz} and 
Fig.~\ref{fig-helix-pressure} shows the bolus transport for this case. A high pressure region exists along the 
contraction segment, which pushes the bolus down, similar to what is observed in Case 1 in Section~\ref{sec_case_axial_mucosal}. 
The details of deformation are illustrated in Fig.~\ref{fig-helix-csa}, which also shows the four distinct stages. 
However, the deformation pattern is more regular than the previous cases, which may be attributed to the helical 
configuration of the mucosal layer. As Fig. 17 shows, high pressure and increased muscle CSA overlaps, 
which indicates a synchrony between CM contraction and LM shortening. However, no distinct pressure peak exists, which 
is different from results of Case 1 in Section~\ref{sec_case_axial_mucosal}. This is probably because the helical mucosal layer  
is less ``squeezed", which is evidenced by the slightly decreased CSA.

\section{Remarks on the fiber-based immersed boundary method}
\label{sec_limitations_IBM}

In the previous three sub-sections, we showed results for esophageal transport with new insights including roles of mucosal layer and information on lumen pressure and kinematics resulting from the synchrony between CM contraction and LM shortening, in 
spite of the complexity of the configuration. Here, we add remarks on the application 
of the classical fiber-based IB method~\cite{CSPeskin02} to problems involving significant 
deformations. First, it is noted that the choice of number of 
Lagrangian points per Eulerian grid is challenging. Two Lagrangian grid points per Eulerian grid are 
recommended in each direction to stop the fluid from leaking through the immersed structure~\cite{CSPeskin02}. 
However, the esophagus dilates significantly during the bolus transport. It is important to ensure that there 
are at least two grid points per Eulerian grid in each direction in the \emph{dilated} state. This leads to a 
practical requirement that there should be more than two Lagrangian points per Eulerian grid 
in each direction in the \emph{rest} and \emph{relaxation} state. Increasing Lagrangian grid refinement relative to the Eulerian grid does not necessarily imply higher 
accuracy, but instead over constrains the system. Hence, we need to verify that the results are 
reasonable (Fig.~\ref{fig-CSA_pressure}). The fact that the number of Lagrangian grid points are determined 
via simulation trials is an undesirable feature. 

One remedy, not within the scope of this work, would be to use an adaptive Eulerian mesh, where the adaptive 
criterion is based on Lagrangian grid spacing. Current adaptive meshing of the Eulerian grid is based on the magnitude of 
velocity gradients. Another approach could be to have an adaptive Lagrangian discretization such that the number of 
Lagrangian grid points in the rest and dilated states would differ if the Eulerian grid size is fixed.

The second issue is related to spurious deformations at the scale of the Lagrangian grid. Multiple Lagrangian
grid points per Eulerian grid can result in spurious deformation modes at the Lagrangian grid scale. 
These modes are internal to the Lagrangian grid and thus not resolved by the fluid solution which is at 
the scale of the Eulerian grid. As a result, in case of esophageal transport, the relaxed configuration 
recovered after a bolus has passed, has residual deformations on the Lagrangian grid scale. These deformations are 
pronounced in the compliant mucosal layer.

The third issue pertains to spurious deformations leading to errors in the incompressibility of the Lagrangian
structure. The incompressibility constraint is imposed on the Eulerian grid scale. This is not strictly sufficient to impose incompressibility
on the Lagrangian grid which is at a sub-grid scale with respect to the Eulerian grid. 

One remedy for both issues two and three is to use a spring configuration with diagonal springs in addition
to springs oriented in the three orthogonal directions or to use a helical fiber configuration. Such an approach has been used in prior studies involving swimming of two-dimensional eels~\cite{Bhalla2013446,Tytell29102010}. Another remedy for these issues 
could be to use finite element based Lagrangian immersed structure instead of a fiber-based structure~\cite{Griffith2012IBFE}. 
This is being pursued by us but it is not within the scope of this work.

\section{Conclusions}

In this work, we introduce a method based on the volumetric patch to characterize the elasticity of  the immersed fiber-based structure employed in the typical IB method~\cite{CSPeskin02,Griffith2012,Griffith2009}. Model verification is performed via comparisons between the computational results and an analytic solution to an idealized tube dilation problem. Low relative errors (5\%) are obtained. To study the esophageal transport, we develop a fully resolved active musculo-mechanical model that is able to incorporate the liquid bolus, multi-layered esophageal wall, and muscle activation into a unified model. We present
three cases of the esophageal transport that differ in the choice of muscle activation model and mucosal fiber arrangement, thereby demonstrating the capabilities and generality of the model. The key feature of bolus transport observed experimentally is a ``tear-drop" bolus driven by a muscle contraction wave. This is captured in our simulations. Moreover, new insights are also provided by fully resolved simulations. The simulations show that perfect synchrony between LM shortening and CM contraction leads to an overlap of the high pressure region and an increased muscle CSA. This helps to relate clinical test data from manometry and ultrasound image to the underlying neurally-controlled activity, such as the coordination between CM contraction and LM shortening. Detailed information on kinematics elucidates the role of the mucosal layer in shaping the bolus. Specifically, the mucosal layer provides distensibility to the esophageal lumen, and lubricates the running bolus. We remark that this is the first study that directly looks at the interaction between the bolus and the mucosal layer. 
Future work should include a parametric study of the effect of changing mucosal property on bolus transport. This will help to understand certain esophageal diseases that are related to the inflammation of the mucosa.

\section*{Acknowledgements} 

The support of grant R01 DK079902 (JEP) and R01 DK56033 (PJK) from the National Institutes of Health, USA is gratefully acknowledged. B.E.G.~acknowledges research support from the  National Science Foundation (NSF awards DMS-1016554 and ACI-1047734).

\section*{Appendix A: Analysis of a fiber-based tube dilation problem} \label{appendix_derivation}
Here we derive eq.~\eqref{eqn_P_Ur} for the tube dilation problem presented in 
Section~\ref{sec_tube_dilation}. The governing equations for an incompressible fluid are

\begin{align}
\rho\left(\D{\u}{t} + \u \cdot \grad \u \right) & = \grad \cdot \vec{\sigma}_\text{f}  = -\grad p + \mu \lap \u \label{eqn_Appendix_momentum} \\
\grad \cdot \u & = 0,
\end{align}
in which $\sigma_\text{f}$ is the fluid stress, $\u$ is the fluid velocity, $\rho$ is the fluid density, $\mu$ is the fluid 
viscosity and $p$ is the pressure imposing the incompressibility constraint. For a 
neutrally buoyant incompressible viscoelastic structure the governing equations are

\begin{align}
\rho\left(\D{\u}{t} + \u \cdot \grad \u \right) &= \grad \cdot \vec{\sigma}_\text{e} = -\grad p_\text{e} + \grad \cdot \bar{\vec{\sigma}} + \mu \lap \u \label{eqn_Appendix_reduce_momentum} \\
\grad \cdot \u & = 0, \label{eqn_Appendix_continuity} 
\end{align}
in which $\vec{\sigma}_\text{e}$ is the stress tensor in the elastic structure, $p_\text{e}$ is the pressure that imposes the incompressibility constraint in the elastic structure, $\bar{\vec{\sigma}}$ is the unknown deviatoric part of the elasticity tensor. The continuity of traction and velocity at the fluid-solid interface
implies

\begin{equation}
(-p\mathbb{I} + \mu \grad \u)\cdot \vec{n} = (- p_\text{e}\mathbb{I} + \bar{\vec{\sigma}} + \mu \grad \u) \cdot \vec{n}, \label{app_stress_balance}
\end{equation}
in which $\vec{n}$ denotes the outward normal vector to the structure interface. The above equation implies that there is generally a pressure discontinuity at the fluid-structure interface. For the dilation problem at steady state, eq.~\eqref{eqn_Appendix_momentum} implies a uniform pressure in the tube, denoted as $P_\text{inner}$, since the inertial and viscous terms vanish. Thus, eq.~\eqref{app_stress_balance} becomes

\begin{equation}
 P_\text{inner} = \vec{n} \cdot (p\mathbb{I})\cdot \vec{n} = \vec{n} \cdot ( p_\text{e} \mathbb{I} - \bar{\vec{\sigma}}) \cdot \vec{n}. \label{app_stress_balance_steady_state}
\end{equation}

For the fiber-based structure, let $R$ and $r$ denote the initial and deformed radial coordinates, respectively, at each material point. At the boundaries of the tube, let $R_\text{i}$ and $R_\text{o}$ denote the initial inner and outer radius, and $r_\text{i}$ and $r_\text{o}$ denote the current inner and outer radius, respectively. Then we can obtain the relationship: $R = R(r)$, based on volume conservation and plain-strain deformation:

\begin{equation}
 R(r) = \sqrt{r^2 - r_\text{i}^2 + R_\text{i}^2}  \label{eq_R}
\end{equation}
The elasticity of the fiber-based tube is represented by three families of springs: circumferential, radial, 
and axial springs. In the $(r,\theta)$~plane, only deviatoric stresses along the radial and circumferential orientations are nonzero,

\begin{align}
\bar{\sigma}_{rr} &= S_r \left(1-\frac{dR}{dr}\right)  \label{eqn_Appendix_fr} \\
\bar{\sigma}_{\theta\theta} &= S_\theta \frac{r-R}{r}. \label{eqn_Appendix_ft}  
\end{align}
Here, $S_r$ and $S_\theta$ are the moduli of radial and circumferential fibers, respectively. At steady state, the inertial term in the momentum equations of fluid and structure vanishes, and we have $\grad \cdot \vec{\sigma}_\text{e} = 0$. Therefore, along the $r$-direction we get

\begin{equation}
\frac{d\sigma_{\text{e},rr}}{dr} + \frac{\sigma_{\text{e},rr} - \sigma_{\text{e},\theta\theta}}{r} = \frac{d(\bar{\sigma}_{rr}-p_\text{e} )}{dr} + \frac{\bar{\sigma}_{rr} - \bar{\sigma}_{\theta \theta}}{r} = 0. \label{eq_r_balance}
\end{equation}
This gives us the equation for $p_\text{e}$ as

\begin{equation}
\frac{dp_\text{e}}{dr} = \frac{d\bar{\sigma}_{rr}}{dr} + \frac{\bar{\sigma}_{rr} - \bar{\sigma}_{\theta \theta}}{r} 
= S_r \left(\frac{1}{r} -\frac{1}{r}\frac{dR}{dr} - \frac{d^2R}{dr^2}\right) - S_\theta\left(\frac{1}{r} - \frac{R}{r^2}\right). \label{eq_p_e}
\end{equation}
We impose zero-traction boundary condition on the four lateral surfaces of the fluid domain. At steady state, this implies that the exterior fluid pressure is zero, so that traction continuity at the exterior fluid-solid interface implies,

\begin{equation}
0 = \vec{n} \cdot (- p_\text{e} \mathbb{I} + \bar{\vec{\sigma}}) \cdot \vec{n} = -p_\text{e}(r_\text{o}) + \bar{\sigma}_{rr}(r_\text{o}). \label{eqn_Appendix_bc2}
\end{equation}
The solution to eq.~\eqref{eq_p_e} for $p_\text{e}$ with the boundary condition given by 
eq.~\eqref{eqn_Appendix_bc2} reads as

\begin{equation}
p_\text{e}(x)=S_r\left[1-\frac{dR}{dr}(r_\text{o})\right] +  \int_{r_\text{o}}^{x} S_r\left(\frac{1}{r} -\frac{1}{r}\frac{dR}{dr} - \frac{d^2R}{dr^2}\right) - S_\theta\left(\frac{1}{r} - \frac{R}{r^2}\right)\; \text{dr} \label{eqn_Appendix_pr}
\end{equation}
The inner pressure, $P_\text{inner}$ is obtained from eq.~\eqref{app_stress_balance_steady_state} as

\begin{align}
P_\text{inner} &= p_\text{e}(r_\text{i}) -S_r\left[1-\frac{dR}{dr}(r_\text{i})\right] \nonumber \\ 
&= S_r\left[\frac{dR}{dr}(r_\text{i})-\frac{dR}{dr}(r_\text{o})\right] +  \int_{r_\text{o}}^{r_\text{i}} S_r\left(\frac{1}{r} -\frac{1}{r}\frac{dR}{dr} - \frac{d^2R}{dr^2}\right) - S_\theta\left(\frac{1}{r} - \frac{R}{r^2}\right)\; \text{dr}.
 \label{eqn_Appendix_Pinner_FE}
\end{align}
Under dilation and in the absence of shearing motions, the radial fibers will become compressed. For compression-resistant fibers such as those used in this model, this is an unstable configuration, and the minimum energy configuration is obtained when the layers of the tube rotate to release the energy of compression. In this configuration, the radial fibers do not contribute to the elastic stress. Therefore, by letting $S_\theta = S$ and $S_r = 0$, the fluid pressure at the inner surface $r_i$ becomes 

\begin{equation}
P_\text{inner} =  \int_{r_\text{o}}^{r_\text{i}}  - S\left(\frac{1}{r} - \frac{R}{r^2}\right)\; \text{dr}. \label{eq_Appendx_PInner_int}
\end{equation}
Substituting eq.~\eqref{eq_R} into eq.~\eqref{eq_Appendx_PInner_int}, we obtain the explicit form of  $P_\text{inner}$ as

\begin{align}
 P_\text{inner} &=S\left[\log(r_\text{i} + \sqrt{r_\text{i}^2 -C}) - \frac{\sqrt{r_\text{i}^2 - C}}{r_\text{i}}\right] \\ \nonumber
                & - S\left[\log \left(r_\text{o} + \sqrt{r_\text{o}^2 -C}\right) + \frac{\sqrt{r_\text{o}^2 - C}}{r_\text{o}} - \log\left(\frac{r_\text{i}}{r_\text{o}}\right)\right]. 
\label{eq_PInner_expression}
\end{align}
Here  $C = r_\text{i}^2 - R_\text{i}^2 $, $r_\text{i} = R_\text{i} + U(r_\text{i})$, $U(r_\text{i})$ denotes the radial displacement of the inner surface of the tube, and $r_\text{o} = \sqrt{R_\text{o}^2 - R_\text{i}^2 + r_\text{i}^2} $.

%\section*{References}

\begin{flushleft}
  \bibliography{EsoDraft}
\end{flushleft}

%% put all the figures here for review
\newpage

 % -------- model geometry

% ------------------ uniform case

% ---------- non-uniform case

% -------------------------------- helical case

\begin{figure}[ht]
 \centering
 \includegraphics[scale = 0.35]{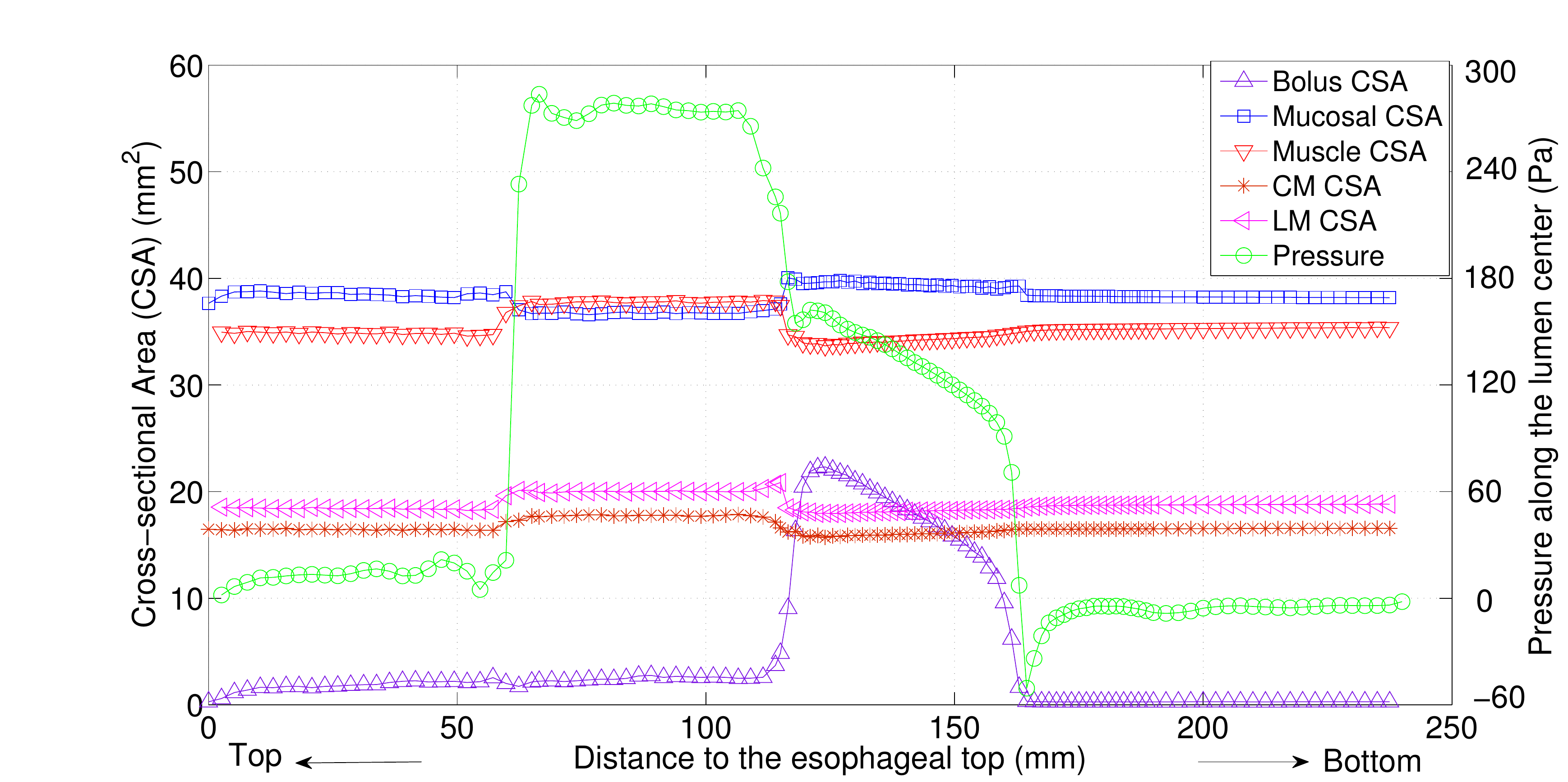}
  \caption{The cross-sectional area (CSA) of the bolus and the esophageal components, and the lumen 
 pressure along its central line: $x=0,y=0$, at $t=1.2$ s for Case 3 in Section~\ref{sec_case_helical_mucosal}. }
 \label{fig-CSA_pressure_helical}
\end{figure}

\end{document}